\documentclass[twocolumn]{aastex631}

\usepackage{multirow}
\usepackage{graphicx}
\usepackage{CJK}
\usepackage[figuresright]{rotating}
\usepackage{longtable}
\usepackage{tabularx}

\begin{document}
\begin{CJK*}{UTF8}{gbsn}
\title{Submillimeter and Mid-Infrared Variability of Young Stellar Objects in the M\,17 \ion{H}{2} Region}

\author[0000-0003-0849-0692]{Zhiwei Chen (陈志维)}
\affiliation{Purple Mountain Observatory, Chinese Academy of Sciences \\
10 Yuanhua Road, 210023 Nanjing, China}

\author[0000-0002-6773-459X]{Doug Johnstone}
\affiliation{NRC Herzberg Astronomy and Astrophysics, 5071 West Saanich Rd, Victoria, BC, V9E 2E7, Canada}
\affiliation{Department of Physics and Astronomy, University of Victoria, Victoria, BC, V8P 5C2, Canada}
 
\author[0000-0003-1894-1880]{Carlos Contreras Pe\~{n}a}
\affiliation{Department of Physics and Astronomy, Seoul National University, 1 Gwanak-ro, Gwanak-gu, Seoul 08826, Korea}
\affiliation{Research Institute of Basic Sciences, Seoul National University, Seoul 08826, Republic of Korea}
 
\author[0000-0003-3119-2087]{Jeong-Eun Lee}
\affiliation{Department of Physics and Astronomy, Seoul National University, 1 Gwanak-ro, Gwanak-gu, Seoul 08826, Korea}
\affiliation{SNU Astronomy Research Center, Seoul National University, 1 Gwanak-ro, Gwanak-gu, Seoul 08826, Korea}

\author{Sheng-Yuan Liu}
\affiliation{Academia Sinica Institute of Astronomy and Astrophysics, 11F of AS/NTU Astronomy-Mathematics Building, No.1, Sec. 4, Roosevelt Rd, Taipei 10617, Taiwan, R.O.C.}

\author[0000-0002-7154-6065]{Gregory Herczeg (沈雷歌)}
\affiliation{Kavli Institute for Astronomy and Astrophysics, Peking University, Yiheyuan Lu 5, Haidian Qu, 100871 Beijing, Peoples Republic of China}
\affiliation{Department of Astronomy, Peking University, Yiheyuan 5, Haidian Qu, 100871 Beijing, China}

\author[0000-0002-6956-0730]{Steve Mairs}
\affiliation{NRC Herzberg Astronomy and Astrophysics, 5071 West Saanich Rd, Victoria, BC, V9E 2E7, Canada}
\affiliation{East Asian Observatory, 660 N. A`oh\={o}k\={u} Place,
Hilo, Hawai`i, 96720, USA}

\author[0000-0001-8467-3736]{Geumsook Park}
\affiliation{Telepix Co., Ltd., 17, Techno 4-ro, Yuseong-gu, Daejeon 34013, Republic of Korea}
\affiliation{Research Institute of Natural Sciences, Chungnam National University, 99 Daehak-ro, Yuseong-gu, Daejeon 34134, Republic of Korea}
\affiliation{Korea Astronomy and Space Science Institute, 776 Daedeokdae-ro, Yuseong-gu, Daejeon 34055, Republic of Korea}

 
\author[0000-0003-2412-7092]{Kee-Tae Kim}
\affiliation{Korea Astronomy and Space Science Institute, 776 Daedeokdae-ro, Yuseong-gu, Daejeon 34055, Republic of Korea}
\affiliation{University of Science and Technology, Korea (UST), 217 Gajeong-ro, Yuseong-gu, Daejeon 34113, Republic of Korea}
 
\author{Mi-Ryang Kim}
\affiliation{Department of Physics and Astronomy, Seoul National University, 1 Gwanak-ro, Gwanak-gu, Seoul 08826, Korea}

\author{Keping Qiu}
\affiliation{School of Astronomy and Space Science, Nanjing University Xianlin Campus, 163 Xianlin Avenue, Qixia District, Nanjing, Jiangsu, China, 210023}
 
\author{Yao-Te Wang}
\affiliation{Academia Sinica Institute of Astronomy and Astrophysics, 11F of AS/NTU Astronomy-Mathematics Building, No.1, Sec. 4, Roosevelt Rd, Taipei 10617, Taiwan, R.O.C.}
\affiliation{Graduate Institute of Astrophysics, National Taiwan University, No. 1, Sec. 4, Roosevelt Rd., Taipei 10617, Taiwan, R.O.C.}
 
\author{Xu Zhang}
\affiliation{School of Astronomy and Space Science, Nanjing University Xianlin Campus, 163 Xianlin Avenue, Qixia District, 210023 Nanjing, China}

\author{Megan Reiter}
\affiliation{Department of Physics and Astronomy, Rice University, 6100 Main St - MS 108, Houston, TX 77005, USA}
\author{The JCMT Transient Team}








\begin{abstract}
We conducted a comprehensive analysis of young stellar object (YSO) variability at submillimeter and mid-infrared (mid-IR) wavelengths for the M\,17 \ion{H}{2} region, using 3.5 years monitoring data from the JCMT Transient Survey at $450$ and $850\,\mu$m and 9 years mid-IR monitoring data from the NEOWISE mission. Our study encompasses observations of 198 and 164 bright submillimeter peaks identified within the deep JCMT coadded maps at 450 and $850\,\mu$m, and 66 YSOs seen by NEOWISE W2 that were previously identified in mid-IR observations. We find one robust linear submillimeter variable, an intermediate mass protostar, with a $4\%$ peak flux change in 3.5 years of JCMT monitoring that sets a lower limit of $16\%$ luminosity increase for the source. At mid-IR wavelengths, our analysis reveals secular and stochastic variability in 22 YSOs, with the highest fraction of secular variability occurring at the earliest evolutionary stage. This mid-IR fractional variability as a function of evolutionary stage result is similar to what has previously been found for YSO variability within the Gould Belt and the intermediate-mass star formation region M17\,SWex, though overall less variability is detected in M\,17 in submillimeter and mid-IR. We suspect that this lower detection of YSO variability is due to both the greater distance to M\,17 and the strong feedback from the \ion{H}{2} region. Our findings showcase the utility of multiwavelength observations to better capture the complex variability phenomena inherent to star formation processes and demonstrate the importance of years-long monitoring of a diverse selection of star-forming environments.

\end{abstract}



\section{Introduction} \label{sect:intro}
Episodic accretion events of low-mass young stellar objects (YSOs) have been observed for decades, primarily through optical outbursts
\citep[see review by][]{2014prpl.conf..387A,2023ASPC..534..355F}.
Recent discoveries at longer wavelengths have included YSO bursts at younger stages of protostellar evolution
\citep{2007A&A...470..211K,2012AJ....144..192M,2015ApJ...800L...5S,2017MNRAS.465.3011C,2019AJ....158..241H,2021ApJ...920..132P,2021ApJ...920..119L}, closer to the peak time of the stellar mass assembly. 
However, only a few bursts have been identified from massive YSOs, usually uncovered by monitoring maser emission with confirmation through infrared follow-up observations \citep{2017NatPh..13..276C,2020NatAs...4..506B,2021A&A...646A.161S,2023NatAs...7..557B}.

In the earliest stages of star formation, numerical simulations predict that episodic accretion bursts might contribute a substantial part to the final stellar mass \citep[e.g.][]{2015ApJ...805..115V,2019MNRAS.482.5459M}. 
Episodic accretion was proposed by \citet{1990AJ.....99..869K} as a solution to the ``protostellar luminosity problem" observed in Class I YSOs. At the time, these sources appeared to be underluminous on average compared to expectations based on simple accretion arguments. 
Updated measurements of protostellar luminosities and lifetimes have modified this historical perspective into the ``protostellar luminosity spread problem", the scatter observed around the best-fit isochrones in pre-main-sequence stellar clusters \citep{2009ApJ...702L..27B,2012ApJ...756..118B,2017A&A...599A..49K}. Time-dependent accretion may naturally explain this wide range of protostellar luminosities that span $3-4$ orders of magnitude \citep[see review by][]{2023ASPC..534..355F}. The higher frequency of accretion bursts during the earliest stages of YSOs, as indicated 
by the statistics of variable YSOs in the Gould Belt \citep{2021ApJ...920..132P,2022ApJ...924L..23Z}, agrees with the expectation of theoretical models of angular momentum transport in accretion disks \citep[e.g.][]{2014ApJ...795...61B,2015ApJ...805..115V}. However, YSOs in the early stages are still deeply embedded in their nascent, dusty envelopes and are thus too heavily extincted for an accretion burst to be directly observed at optical or near-infrared wavelengths. This radiation, emitted by the deeply embedded YSO, is absorbed by the dense envelope and re-radiated at longer wavelengths, with the emitted spectral energy distribution from the envelope peaking at mid- to far-infrared wavelengths. Subsequent changes in observed flux at these, and longer, wavelengths trace the heating and cooling of the envelope due to the internal accretion-driven luminosity variability \citep{2013ApJ...765..133J,2019MNRAS.487.4465M,2019MNRAS.487.5106M,2020ApJ...895...27B}.

In principle, the best wavelength range to monitor the accretion variability of deeply embedded protostars is 20--200 $\mu$m
\citep{2019MNRAS.487.4465M,2024AJ....167...82F}, but this window is difficult to access with ground-based telescopes.  At longer wavelengths, submillimeter monitoring programs have begun to reveal protostellar variability, although at modest levels \citep[e.g.][]{2021ApJ...920..119L,2022ApJ...937...29F,2024ApJ...966..215M}.  At shorter wavelengths, monitoring programs with Spitzer and WISE have detected mid-IR variability of YSOs up to several magnitudes \citep{2011ApJ...731...53M,2013MNRAS.430.2910S,2018AJ....155...99W,2021ApJ...920..132P,2024ApJ...962...38L}. Although mid-IR variability can also be affected by changes in the extinction and geometry of asymmetric disks \citep[e.g.][]{2015MNRAS.451...26S,2019A&A...625A..45N,2021AJ....161...61C}, in the few sources monitored for both submillimeter and mid-IR variability, the correlated changes are consistent with expectations of accretion variability \citep{2020MNRAS.495.3614C}.

Most previous results focus on nearby low-mass protostars. The variability properties towards the higher mass end are much less explored, with a very limited sample of accretion-bursting MYSOs \citep[e.g.][]{2021ApJ...922...90C,2024A&A...688A...8W}. All of these accretion-bursting MYSOs exhibit contemporary variability in infrared and submillimeter luminosity and in maser emission \citep{2017NatPh..13..276C,2017ApJ...837L..29H,2018ApJ...863L..12L,2020NatAs...4.1170C,2020NatAs...4..506B,2021A&A...646A.161S,2021ApJ...912L..17H,2021ApJ...922...90C,2022PASJ...74.1234H}. The question remains whether similar accretion-driven luminosity bursts to those seen for lower mass YSOs occur in MYSOs 
and whether these processes might differ in frequency, amplitude, and duration as a function of protostellar mass \citep{2023ASPC..534..355F}. 




The JCMT Transient survey originally focused exclusively on nearby ($<500$ pc) low-mass star-forming regions and expanded in 2020 to monitor six additional fields towards regions of intermediate- to high-mass star formation \citep{2017ApJ...849...43H,2024ApJ...966..215M}.
These six fields are at distances of a few kiloparsec, at which the JCMT beams at 450 and $850\,\mu$m are still capable of resolving sub-parsec dust condensations. Some of these fields were selected because they had been previously found to host variable YSOs showing evidence of accretion bursts \citep{2018ApJ...863L..12L,2019ApJS..242...27P,2021ApJ...922...90C,2022ApJ...930..114W}. Two fields are from the M\,17 complex \citep{2020ApJ...891...66N}, the well-known Galactic \ion{H}{2} region M\,17 \citep{2008ApJ...686..310H,2012PASJ...64..110C,2020ApJ...888...98L} and the filamentary molecular cloud extended to southwest of the M\,17 \ion{H}{2} region \citep[M17\,SWex;][]{2010ApJ...714L.285P}. The analyses of the submillimeter and mid-IR variability of YSOs in M17\,SWex are presented in \citet{2024AJ....168..122P}.

\begin{figure*}
\centering
\includegraphics[width=0.7\textwidth]{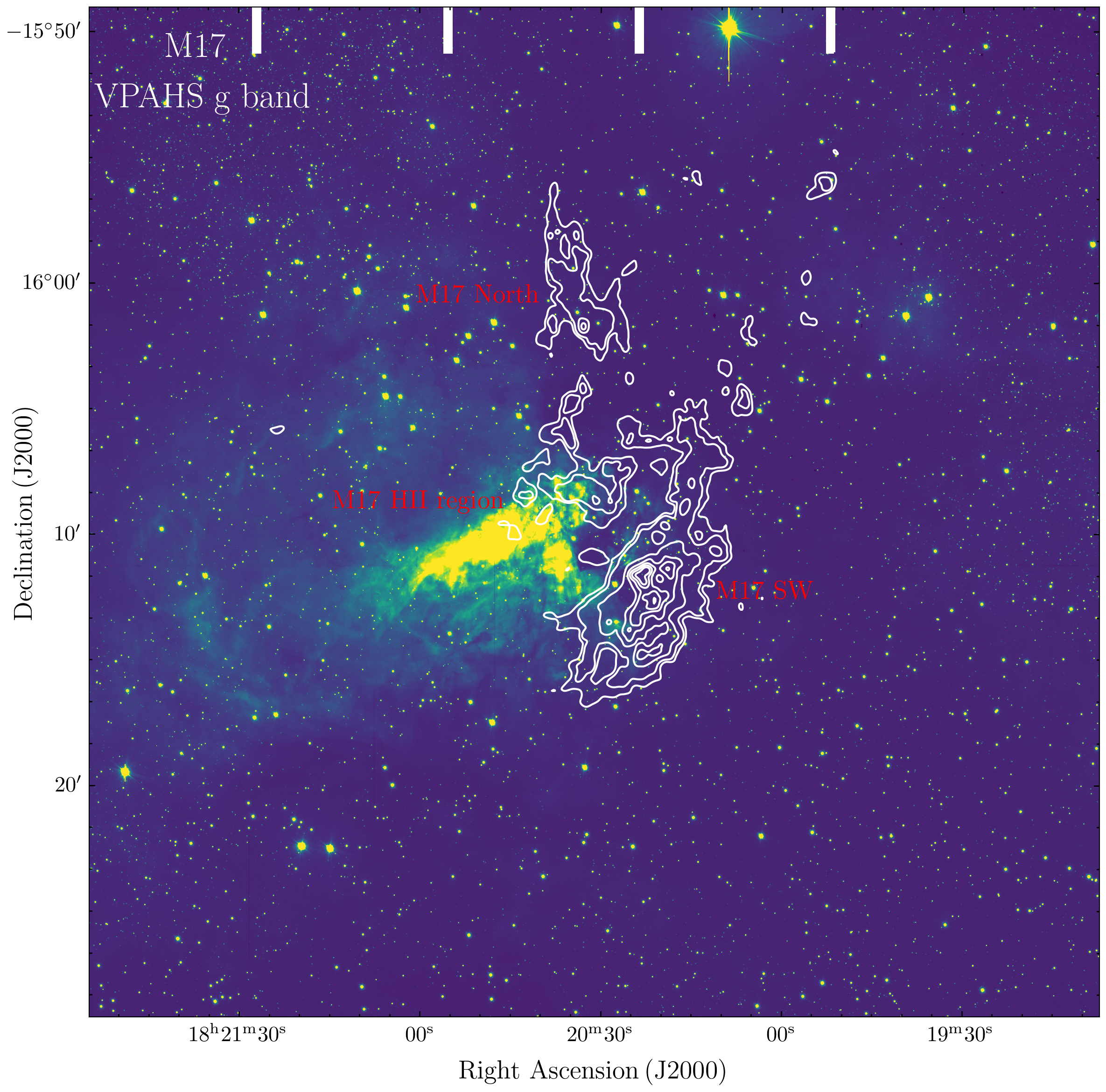}\\
\caption{VPAHS $g$-band image of M\,17, overlaid with $850\,\mu$m dust continuum contours (in levels 0.3, 0.6, 1.8, 3.6, 6.0, 9.0, 1.26, and 1.68 $\mathrm{Jy\,beam^{-1}}$) from JCMT-SCUBA2. The optically bright nebula in M\,17 are labeled as M17 \ion{H}{2} region, while the northern and southwestern clouds are labeled as M17 North and M17 SW, respectively.}
\label{fig:VPAHS_g}
\end{figure*}

The trigonometric distance to the maser sources associated with M\,17 is $\approx2.0$ kpc \citep{2011ApJ...733...25X,2016MNRAS.460.1839C}, which we adopt in this work. This distance is consistent with the distance of $\approx1.7$ kpc measured from \textit{Gaia} astrometry 
\citep{2019ApJ...870...32K,2022A&A...657A.131M,2024A&A...681A..21S}. Figure~\ref{fig:VPAHS_g} shows the SDSS g image of M\,17 from the VST Photometric H$\alpha$ Survey of the Southern Galactic Plane and Bulge \citep[VPHAS;][]{2014MNRAS.440.2036D}. The optically bright nebula outlines the location of the M17 \ion{H}{2} region, with two adjacent clouds in the north (M17 North) and southwest (M17 SW) of the nebula. The massive stars are mostly located within the M17 \ion{H}{2} region \citep{2008ApJ...686..310H}, meanwhile the YSOs are more concentrated towards M17 SW and M17 North \citep{2008ApJ...686..310H,2009ApJ...696.1278P}. Notably, M17~MIR in M17 SW, named due to its visibility only at mid-IR and longer wavelengths, was discovered for its recurrent accretion outbursts \citep{2021ApJ...922...90C,2024ApJ...969L...6Z}. The submillimeter and mid-IR variability of the many other YSOs in M\,17 is still unexplored.

The results of the analyses for the submillimeter and mid-IR variability of YSOs in M\,17 are presented in this paper. The multi-epoch submillimeter and mid-IR data and ancillary data sets are described in Section~\ref{sect:Obs_data}. The analyses of submillimeter and mid-IR variability of YSOs are presented in Section~\ref{sect:sub-mm}. In Section~\ref{sect:discussion}, we focus on the interesting YSOs that are variable at submillimeter or mid-IR wavelengths, and compare the results of M\,17 to those of other fields of the JCMT Transient Survey.


\begin{deluxetable*}{lcccc}
    \centering
    \footnotesize
    \tablecaption{A Summary of JCMT-SCUBA2 Observations at 450 and 850 $\mu$m for M\,17 \label{tab:log}}
\tablehead{
\colhead{ Date} & \colhead{MJD} & \colhead{$\tau\times \mathrm{AM}$ } & \colhead{$\mathrm{RMS_{450}}$} & \colhead{$\mathrm{RMS_{850}}$} \\
\colhead{yyyy-mm-dd} &  &  & \colhead{(mJy beam$^{-1}$)} & \colhead{(mJy beam$^{-1}$)} }
\startdata
2020-02-22 & 58901.7 & 0.05 & 61 & 9 \\
2020-05-21 & 58990.5 & 0.15 & - & 12 \\
2020-06-23 & 59023.5 & 0.14 & - & 13 \\
2020-07-30 & 59060.3 & 0.07 & 116 & 11 \\
2020-09-01 & 59093.2 & 0.05 & 44 & 7 \\
2020-10-10 & 59132.2 & 0.17 & - & 16 \\
2021-03-04 & 59277.7 & 0.06 & 112 & 11 \\
2021-04-06 & 59310.6 & 0.07 & 86 & 9 \\
2021-05-17 & 59351.5 & 0.06 & 135 & 15 \\
2021-06-14 & 59379.3 & 0.09 & 105 & 9 \\
2021-07-22 & 59417.4 & 0.12 & - & 13 \\
2021-08-22 & 59448.2 & 0.12 & 286 & 11 \\
2021-09-27 & 59484.3 & 0.1 & 168 & 10 \\
2021-11-01 & 59519.2 & 0.15 & - & 13 \\
2022-02-19 & 59629.7 & 0.07 & 88 & 11 \\
2022-05-23 & 59722.5 & 0.07 & 67 & 8 \\
2022-06-28\tablenotemark{\scriptsize a}  & 59758.4 & 0.07 & 106 & 13 \\
2022-07-29 & 59789.3 & 0.1 & 190 & 10 \\
2022-08-27 & 59818.2 & 0.07 & 100 & 10 \\
2022-10-01 & 59853.3 & 0.11 & 275 & 12 \\
2023-05-01 & 60065.5 & 0.16 & - & 12 \\
2023-06-06 & 60101.5 & 0.10 & 228 & 11  \\
2023-07-15 & 60140.3 & 0.09 & 201 & 14  \\
2023-08-12 & 60168.3 & 0.06 & 96 & 9  \\
\enddata
\tablenotetext{a}{This epoch is not used for the variability analyses.}
\end{deluxetable*}

\begin{figure*}
\centering
\includegraphics[width=0.995\textwidth]{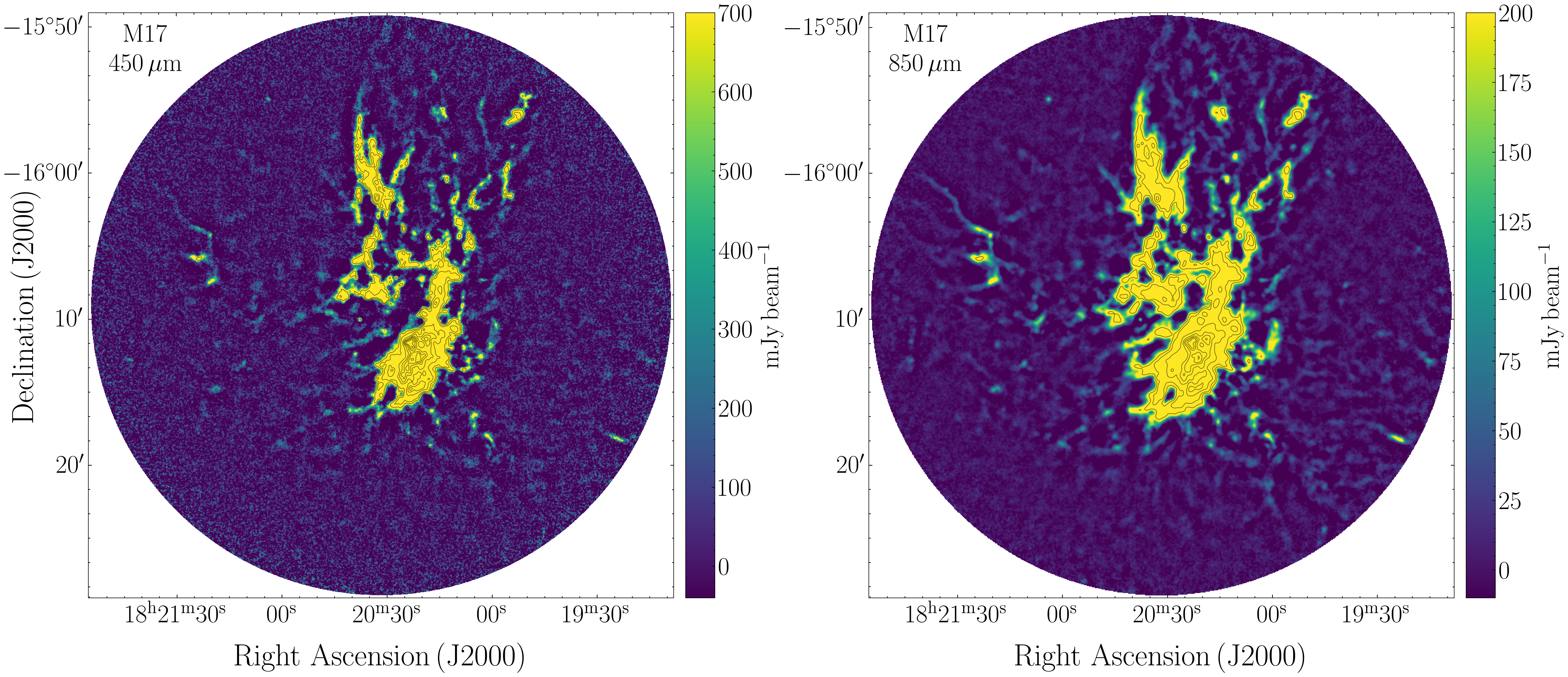}
\caption{JCMT-SCUBA2 co-added 450 (left) and $850\,\mu$m (right) maps for M\,17, overlaid with $450\,\mu$m dust continuum contours (in levels 1, 2, 6, 12, 20, 30, 42, and 56 $\mathrm{Jy\,beam^{-1}}$) and $850\,\mu$m dust continuum contours (in levels 0.3, 0.6, 1.8, 3.6, 6.0, 9.0, 1.26, and 1.68 $\mathrm{Jy\,beam^{-1}}$).}
\label{fig:coadd}
\end{figure*}

\section{Observations and Data}\label{sect:Obs_data}

\subsection{Submillimeter Data Sets}
\subsubsection{JCMT Transient Survey}\label{sect:pipeline}
The JCMT Transient survey is designed to measure submillimeter variability of protostars  \citep{2017ApJ...849...43H} using the Submillimetre Common User Bolometer Array~2 (SCUBA-2) instrument \citep{2013MNRAS.430.2513H} on the James Clerk Maxwell Telescope (JCMT) at the summit of Maunakea, Hawaii. SCUBA-2 observations are performed simultaneously at 450 and $850~\mu$m with effective beam sizes being $10\farcs0\pm0\farcs6$ and $14\farcs4\pm0\farcs3$, or $0.1$ and $0.14$ pc at the distance of M\,17, respectively
\citep{2010SPIE.7740E..1ZK,2021AJ....162..191M}. Our survey uses the \textit{pong} $1800\arcsec$ mapping mode, which scans a field of $30\arcmin$ in diameter and produces a uniform background rms noise within the circular field. 

SCUBA-2 monitoring observations for M\,17 started on February 22, 2020 with a monthly cadence as long as the region is visible by JCMT. During the period from 2020 February to 2023 August, 24 epochs of observations were carried out, although the visit on 28 June 2022 appeared problematic and is excluded from our variability analyses (see Table~\ref{tab:log}). The $850\,\mu$m data have a consistent background rms noise of $\sim12\mathrm{mJy\,beam^{-1}}$ for each of these 23 epochs.  Observations at $450\,\mu$m strongly depend on atmospheric transmission \citep{2024ApJ...966..215M}. In 17 of the 23 epochs, 
the $450\,\mu$m data were obtained with $\tau~ \times~\mathrm{airmass} \leqslant 0.12$ and therefore have a low enough rms for variability analysis.

The data reduction procedure was performed using the iterative map making technique MAKEMAP \citep{2013MNRAS.430.2545C} in the SMURF package within the STARLINK software \citep{2014ASPC..485..391C}. The JCMT Transient Program tweaked the user-defined parameters of MAKEMAP and reduced the survey data with the so-called "R3" version \citep[see][for more details] {2017ApJ...843...55M}. Since the goal of these monitoring observations is to measure the fluxes of individual compact sources over time, the flux and astrometry calibration are important for the detection of the variability of embedded stars at $450$ and $850\,\mu$m.  In addition to the data reduction procedures in above, the JCMT Transient survey developed its own calibration pipelines to align the SCUBA-2 maps with each other \citep{2017ApJ...843...55M,2024ApJ...966..215M}. 
All data from the JCMT Transient survey have been processed with the Pipeline v2, which achieves relative image alignment better than $0\farcs5$, and a relative flux calibration at the 1\% level at $850\,\mu$m and at $<5\%$ at $450\,\mu$m \citep{2024ApJ...966..215M}. 
The ``Relative Flux Calibration Factor" (Relative-FCF; $R-\mathrm{FCF}$) for each epoch delivered with the Pipeline v2 at 450 and $850\,\mu$m are listed in the last two columns in Table~\ref{tab:log}.

\subsubsection{Localized Submillimeter Peaks}\label{sect:submm_source}
The SCUBA-2 maps are co-added to produce the reference maps at $450$ (from 17 epochs) and $850\,\mu$m (from 23 epochs).
%
Figure~\ref{fig:coadd} shows the SCUBA-2 co-added $450$ and $850\,\mu$m maps for M\,17, with contours overlaid to 
highlight the relatively bright submillimeter sources.
The localized submillimeter sources in the co-added $450$ and
$850\,\mu$m maps are identified using the clump identification
algorithm FellWalker \citep{2015A&C....10...22B} in the CUPID
package \citep{2007ASPC..376..425B} found within the STARLINK
software \citep{2014ASPC..485..391C}. Sources are only included
if they are located within a $20\arcmin$ radius of the map
center, in which area rms noises at $450$ and $850\,\mu$m are
nearly uniform. The source catalogues of 448 peaks at $450\,\mu$m
and 584 peaks at $850\,\mu$m are extracted from the co-added
images. Statistical analysis in this paper focuses on the 198
peaks brighter than $0.65\,\mathrm{Jy\,beam^{-1}}$ at $450\,\mu$m
and the 164 peaks brighter than $0.14\,\mathrm{Jy\,beam^{-1}}$ at
850 $\mu$m.
For completeness, Tables~\ref{tab:450_sources} and \ref{tab:850_sources} present the locations, mean peak brightnesses, and variability measurements for all the relatively bright submillimeter sources found by FellWalker at $450$ and $850\,\mu$m, respectively.

With the locations determined, the peak flux of each source in the reference catalogues is measured in each observed epoch to construct initial light curves for all identified objects. The initial light curves are then recalibrated with the $R-\mathrm{FCF}$ in all epochs to generate the calibrated light curves at $450$ and $850\,\mu$m.  

\begin{figure*}
\centering
\includegraphics[width=0.7\textwidth]{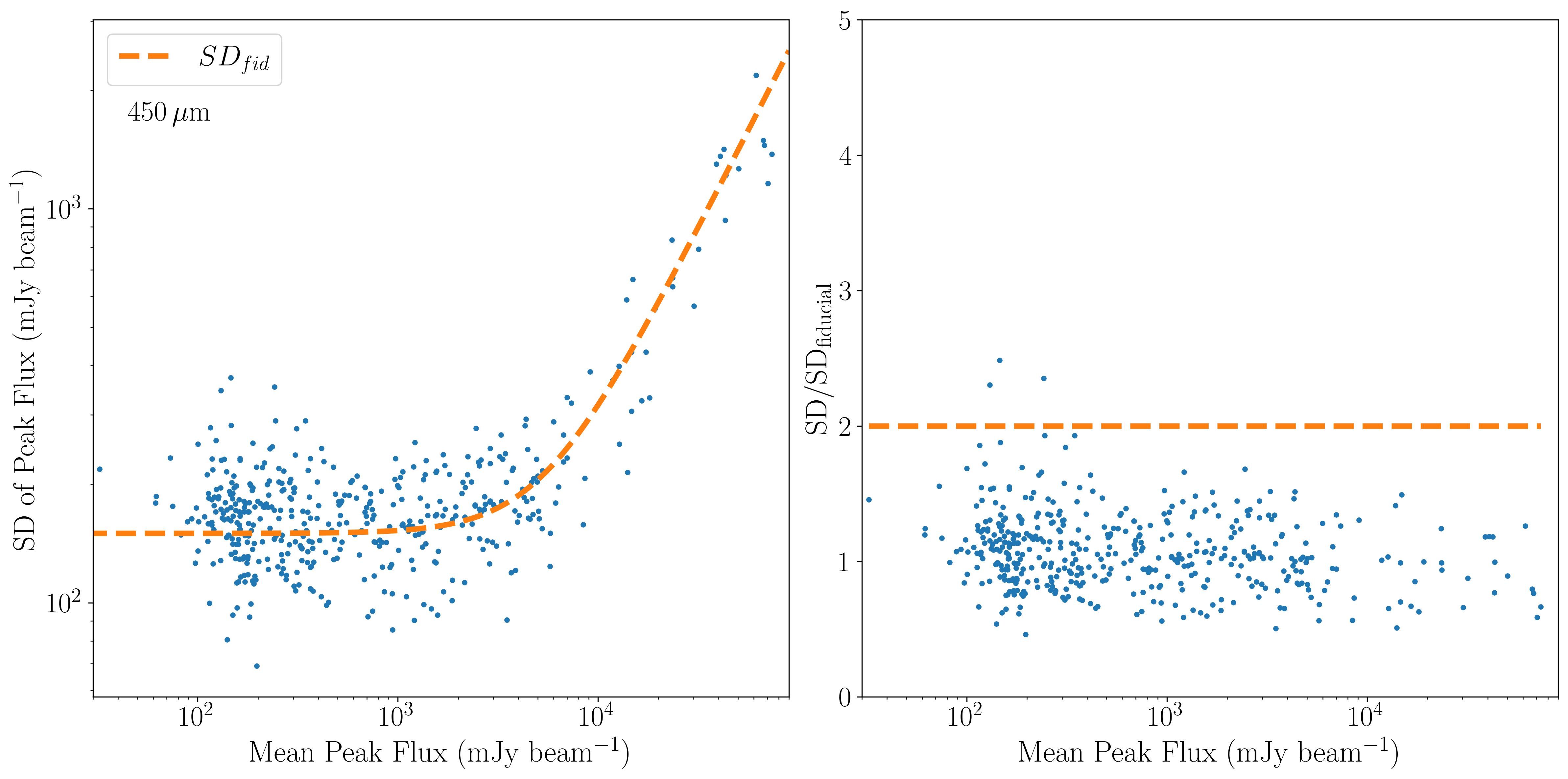}\\
\includegraphics[width=0.7\textwidth]{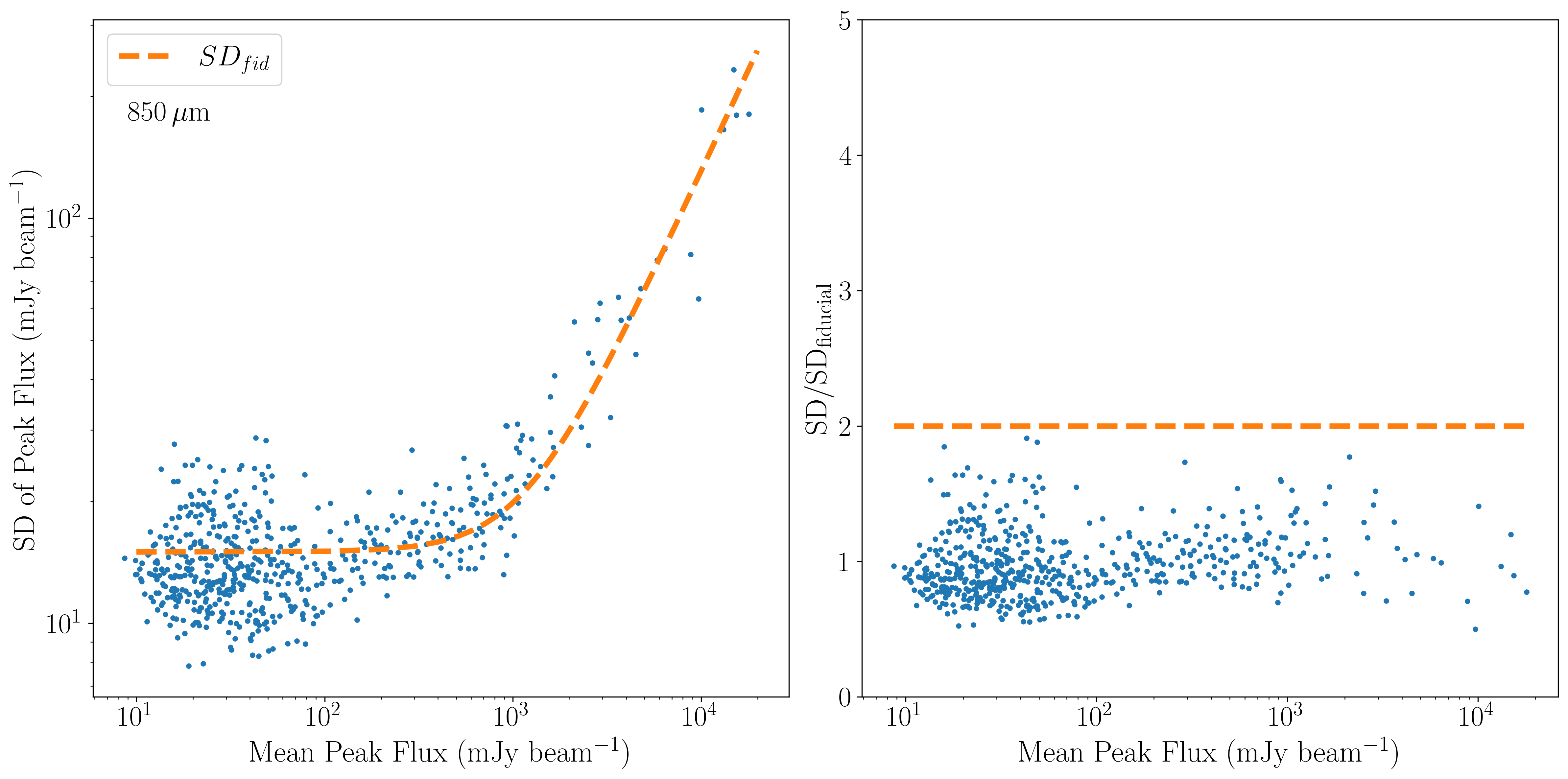}
\caption{Left panels: Scatter plot of measured standard deviation of peak brightness versus measured mean peak brightness for all $450\,\mu$m (top row) and $850\,\mu$m sources (bottom row) in M\,17. The orange dashed curve denotes the fiducial standard deviation, $\mathrm{SD_{fid}}$, for the expected uncertainty in the peak brightness of JCMT-SCUBA2 sources. Right panels: the same as the left panels, but for the normalized standard deviation of peak brightness ($SD/\mathrm{SD_{fid}}$). Here the dashed orange line separates out the potential stochastic variables. }
\label{fig:SD}
\end{figure*}

\subsection{Mid-IR Data Sets}
\subsubsection{Infrared-bright YSOs in M\,17}\label{sect:ysos}
M\,17 is one of the most intensively studied massive star-forming regions in the Galaxy. 
Using data from several infrared studies, we collected a sample of YSOs that are located within the inner $40\arcmin$ diameter area of the JCMT field. The YSOs in the sample are identified via infrared excess from 2MASS and Spitzer observations \citep{2009ApJ...696.1278P,2010ApJ...714L.285P,2013ApJS..209...31P}, with most of the sources being part of the Spitzer/IRAC candidate YSO catalog for the inner galactic midplane \citep[SPICY;][]{2021ApJS..254...33K}. We also include a few massive YSOs classified from SOFIA/FORCAST observations for M17 at $19$ and $37\,\mu$m by \citet{2020ApJ...888...98L}, and an outburst MYSO M 17\,MIR by \citet{2021ApJ...922...90C}. 

Based on the above studies, we compile a catalog of 166 YSOs. These YSOs are predominantly in early evolutionary stages, with 34.9\% Class I objects and 42.8\% Class II objects. Class III YSOs are only a minor fraction (1.2\%) of the sample, likely due to the complexity of identifying them in distant massive star-forming regions. A substantial fraction (21.1\%) of the sample is assigned as `uncertain' objects, whose infrared excess emission can also be explained by the circumstellar envelope surrounding the evolved post-main-sequence stars \citep{2013A&A...557A..51C}.

\subsubsection{Multi-epoch WISE/NEOWISE Data}\label{sect:neowise}
The Wide-field Infrared Survey Explorer (WISE, \citealt{2010AJ....140.1868W}) is a 40\,cm telescope in a low Earth orbit that surveyed the entire sky in 2010 at 3.4, 4.6, 12, and $22\,\mu$m. The angular resolutions in the four bands (W1, W2, W3, and W4) are $6\farcs1$, $6\farcs4$, $6\farcs5$, and $12\farcs0$, respectively. The WISE telescope was reactivated as the Near-Earth Object WISE \citep[NEOWISE][]{2011ApJ...731...53M,2014ApJ...792...30M} program, which used only the short wavelength W1 and W2 bands to search for near-Earth objects. NEOWISE observations of M17 were obtained every 6 months from December 2013 until December 2022. Each visit consists of $10-20$ exposures taken over a few days.

For each source in the YSO sample, we first queried the WISE and NEOWISE single exposure catalogs  \citep{2019ipac.data...I1W,2020ipac.data.I124W} from the NASA/IPAC Infrared Science Archive (IRSA), using a radius $3\arcsec$, the same method used for the previous studies \citep{2020MNRAS.495.3614C,2021ApJ...920..132P,2024AJ....168..122P}. The average values of the Right Ascension (R.A.) and Declination (Decl.) in J2000 epoch are determined from all source positions of the same entry. We then selected single-exposure sources that are located within $2\times sd_d$ of this mean location (where $sd_d$ is the standard deviation of the distances from the mean location). The next step was to group all the single-exposure measurements performed within a few days of each other. 
Because we looked for long-trend mid-IR variability over years, we discarded the brightest and faintest 15\% for each group. Using the remaining 70\% of the group, we estimate the mean modified Julian date (MJD), the mean magnitude in W1 / W2, the mean error and the standard deviation (in magnitude) for each YSO matched in the specific epoch. The measurement error was then calculated by adding, in quadrature, the mean error and standard deviation in each epoch. This method then provides multi-epoch WISE/NEOWISE observations every 6 months. Finally, the WISE/NEOWISE surveys provide up to 19 epochs of W1 and W2 photometry for YSOs in M\,17 in the period from 2010 to 2022. In particular, the multi-epoch W2 magnitudes of one YSO in M\,17,  which is not included by NEOWISE single exposure catalogs, was adopted from our previous paper \citep[M17 MIR;][]{2021ApJ...922...90C}.

To obtain reliable results, we select YSOs with a minimum of 12 epochs in both W1 and W2 and a mean uncertainty $\mathrm{\sigma(mag)}<0.2$ mag. Only 66 of the 166 IR-bright YSOs compiled in \S 2.2.1 meet this requirement. The low number of reliable sources could be explained in part by the lower spatial resolution of \textit{WISE} data as compared with \textit{Spitzer} data, which is used to detect YSOs in our sample (see Section \ref{sect:neowise}). In addition, the strong mid-IR emission from the photodissociation region (PDR) excited by the massive stars in M\,17 worsens the sensitivity of the \textit{WISE} single-exposure maps.

\subsection{Far-IR and millimeter Data Sets}
In order to characterize the far-IR emission from sources, we incorporated publicly available imaging observations from the Herschel Space Observatory (Herschel) and its PACS instrument at 70, 100, and 160\,$\mu$m \citep{2010A&A...518L...2P}. The Herschel data used in this work are Level 3.0 products of the PACS calibration observations towards M\,17 (eight observations with IDs from 1342192767 to 1342192774), provided by the Herschel Science Archive\footnote{http://archives.esac.esa.int/hsa/whs}. The pixel scales are $1\farcs6$, $1\farcs6$, and $3\farcs2$ at 70, 100, and 160\,$\mu$m, and the measured resolutions are $8\farcs0$, $8\farcs0$, and $12\farcs0$, respectively. 

We also include observations by the Atacama Compact Array (ACA) for M\,17, obtained on 2019 April 17 (project ID 2018.1.01091.S, PI: M. Reiter) with 11 antennas of 7 m diameter in a fixed configuration. The projected baselines ranged from $\sim8$ to 48 m. M\,17 was observed in Band 6 (230 GHz, 1.3 mm) with six spectral windows (SPWs), three ($230.4735 - 230.5985$,$220.3335-220.4585$, and $219.4955-219.6205$ GHz) for the CO/$^{13}$CO/C$^{18}$O $J=2-1$ lines and three ($216.998-218.998$, $230.998 - 232.998$, and $230.9345-231.0595$ GHz) for continuum emission. The archival ACA Band 6 data were retrieved from the ALMA Science Archive at the National Radio Astronomy Observatory\footnote{https://almascience.nrao.edu/aq/}. Calibration and imaging were performed with CASA version 5.4 \citep{2007ASPC..376..127M}, using the ALMA pipeline. Only the 1.3 mm continuum data is used in this study. The synthesized beam size of 1.3 mm continuum is $7\farcs2\times4\farcs5$. The continuum sensitivity is $\sim4.2\,\mathrm{mJy\,beam^{-1}}$.


\begin{figure*}
\centering
\includegraphics[width=0.7\textwidth]{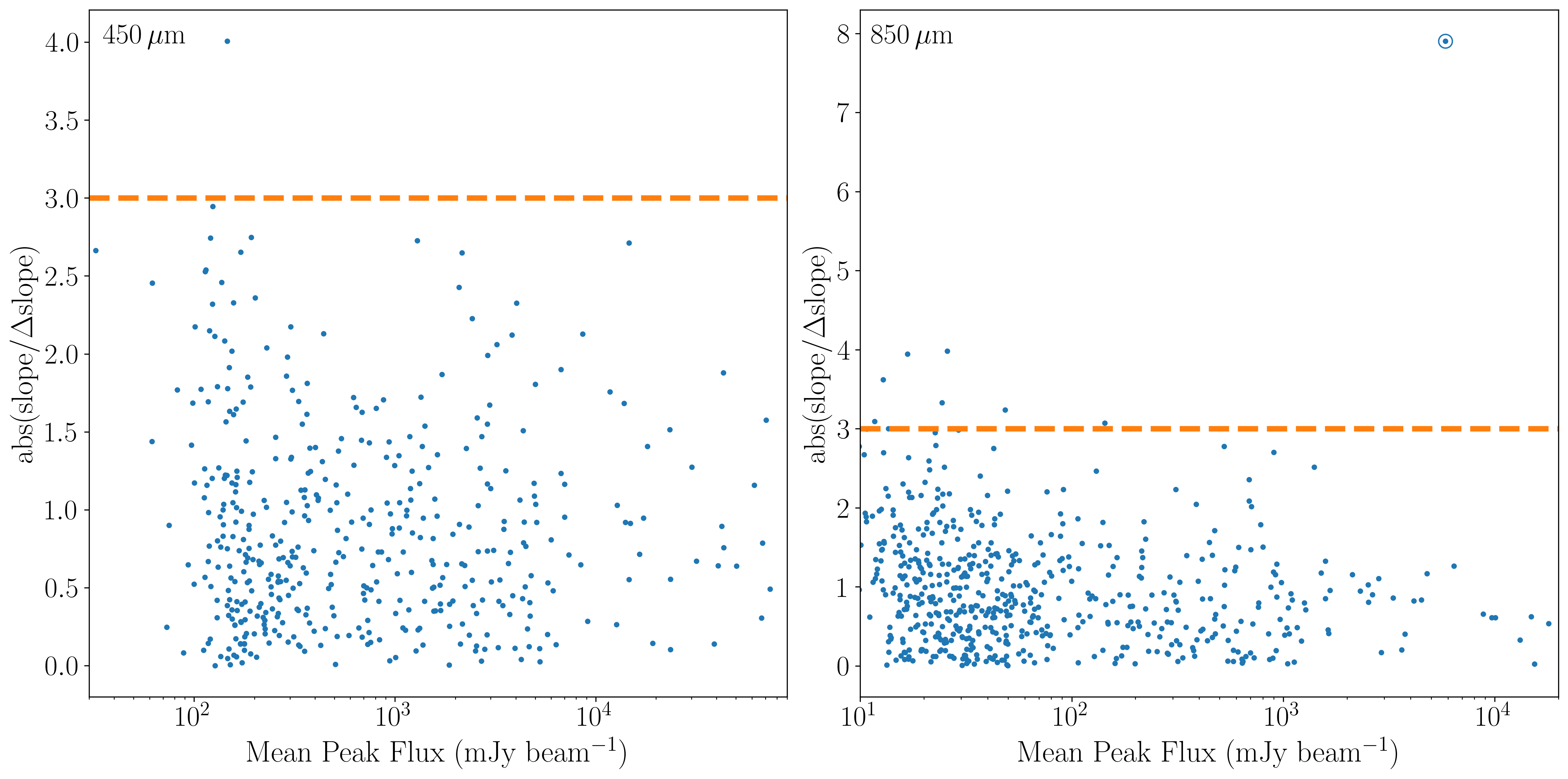}
\caption{The distribution of the absolute value of the best-fit peak brightness slope $S$ ratioed to its uncertainty $\Delta S$ versus mean peak flux for the $450$ (left) and $850\,\mu$m (right) sources in M\,17. The dashed horizontal line in both panels indicates where $|S/\Delta S| = 3$, indicating the threshold for linear variability reliability  as defined by \citet{2018ApJ...854...31J}. Bright $450$ and $850\,\mu$m sources with $|S/\Delta S|>3$ are highlighted by circles. In the $850\,\mu$m plot, only one bright source JCMTPP\,J182023.2 has a value of $|S/\Delta S|$ much higher than the threshold.}
\label{fig:slope}
\end{figure*}

\section{Submillimeter and mid-IR variables} \label{sect:sub-mm}
\subsection{Searching for Submillimeter variables in M\,17}
Submillimeter variability of a given object can be categorized into three types: stochastic, secular, and single-epoch. The stochastic and secular variables can be selected via statistical investigations. We follow the methods outlined by \citet{2018ApJ...854...31J} to search for stochastic and secular variables. For each source, we extract the mean peak brightness in the co-added images and the brightness at the same location in each individual epoch. The standard deviation, $\mathrm{SD}$, in peak brightness across all epochs is then calculated for each source and shown in Figure~\ref{fig:SD}.

The uncertainty in peak brightness for faint sources is dominated by the relatively uniform rms noise per epoch, while for bright sources the uncertainty in the relative calibration of the map will dominate. \citet{2018ApJ...854...31J} proposed a fiducial standard deviation $\mathrm{SD_{fid}}$ for the statistical investigation of the JCMT Transient survey, as 
\begin{equation}
SD_{fid}(i) = \sqrt{(n_\mathrm{RMS})^2 + (\sigma_\mathrm{FCF} \times f_m(i))^2} \,\mathrm{mJy\,beam^{-1}},
\end{equation}
where $n_\mathrm{RMS}$ is the typical rms noise measured across the observed epochs (dominating faint sources), $\sigma_\mathrm{FCF}$ is the expected relative flux calibration uncertainty (dominating bright sources), and $f_m(i)$ is the mean peak flux of source $i$. For the 17 good epochs at $450\,\mu$m, we calculate $n_\mathrm{RMS}\approx 150\,\mathrm{mJy\,beam^{-1}}$ and $\sigma_\mathrm{FCF}\approx2.8\%$, not far off the ensemble values, $130\,\mathrm{mJy\,beam^{-1}}$ and $3.5 - 5\%$, derived for the eight Gould Belt regions recalibrated with Pipeline V2 \citep{2024ApJ...966..215M}.  At $850\,\mu$m, the values of $n_\mathrm{RMS}\approx15\,\mathrm{mJy\,beam^{-1}}$  and $\sigma_\mathrm{FCF}\approx1.3\%$ are very close to the values of $15\,\mathrm{mJy\,beam^{-1}}$ and $1.5\%$ found for M17\,SWex at $850\,\mu$m by \citet{2024AJ....168..122P}.  


The fiducial models for the 450 and $850\,\mu$m sources in M\,17 are plotted as the orange dashed curves in the left column of Figure~\ref{fig:SD}. The majority of  $450$ and $850\,\mu$m sources lie near the fiducial models at each wavelength. To show this result more clearly, the right panels in  Figure~\ref{fig:SD} plot the $\mathrm{SD}$ in units of the fiducial models (normalized $\mathrm{SD}$) as a function of mean source brightness at $450$ and $850\,\mu$m. Sources with a normalized $\mathrm{SD}$ that is clearly higher than unity are potential candidates for submillimeter variables. For example, the normalized $\mathrm{SD}$ of EC\,53 (V371 Ser) was found to be as high as 5.6 at $850\,\mu$m \citep{2018ApJ...854...31J} and clear a strong submillimeter variable \citep{2017ApJ...849...69Y,2020ApJ...903....5L,2022ApJ...937...29F}.

As shown in Figure~\ref{fig:SD}, we find no $450\,\mu$m sources brighter than $0.65\,\mathrm{Jy\,beam^{-1}}$ showing a normalized $\mathrm{SD}$ of peak brightness greater than 2. Also, we find no $850\,\mu$m sources with a normalized $\mathrm{SD}$ of peak brightness greater than 2. From analyses of the standard deviation of the source peak brightness at $450$ and $850\,\mu$m, we find that the submillimeter sources in M17 do not have observable stochastic variability over few year timescales.


We next consider secular peak brightness change over time. We determine the peak brightness slope (fractional change per yr) using a statistical analysis, the same approach undertaken for the eight Gould Belt regions  by \citet{2018ApJ...854...31J}. For each source $i$, we derive a model fit, $f_l(i,t)$, that is linear over time, $t$, with two derived parameters: initial flux, $f_0(i)$ at time $t_0$, the time of the first epoch, and slope, $S(i)$, measured in fractional brightness change per year:
\begin{equation}
f_l(i,t) = f_0(i) \left( 1 + S(i) (t-t_0) \right).
\end{equation}
Furthermore, in order to measure the relevance of any slopes departing from flat ($S=0$; there are no changes in brightness over time) we also compute the uncertainty of the slope, $\Delta S$. In Figure~\ref{fig:slope}, the distribution of the absolute value of the best-fit slope in units of $\Delta S$ versus mean peak brightness is shown for the $450$ and $850\,\mu$m sources in the left and right panels, respectively. We find that none of the $450\,\mu$m sources brighter than $0.65\,\mathrm{Jy\,beam^{-1}}$ have $|S|/\Delta S$ greater than 3, a reasonable threshold for the detection of robust linear submillimeter variables \citep[e.g.][]{2018ApJ...854...31J}.   

\begin{figure}
\centering
\includegraphics[width=0.45\textwidth]{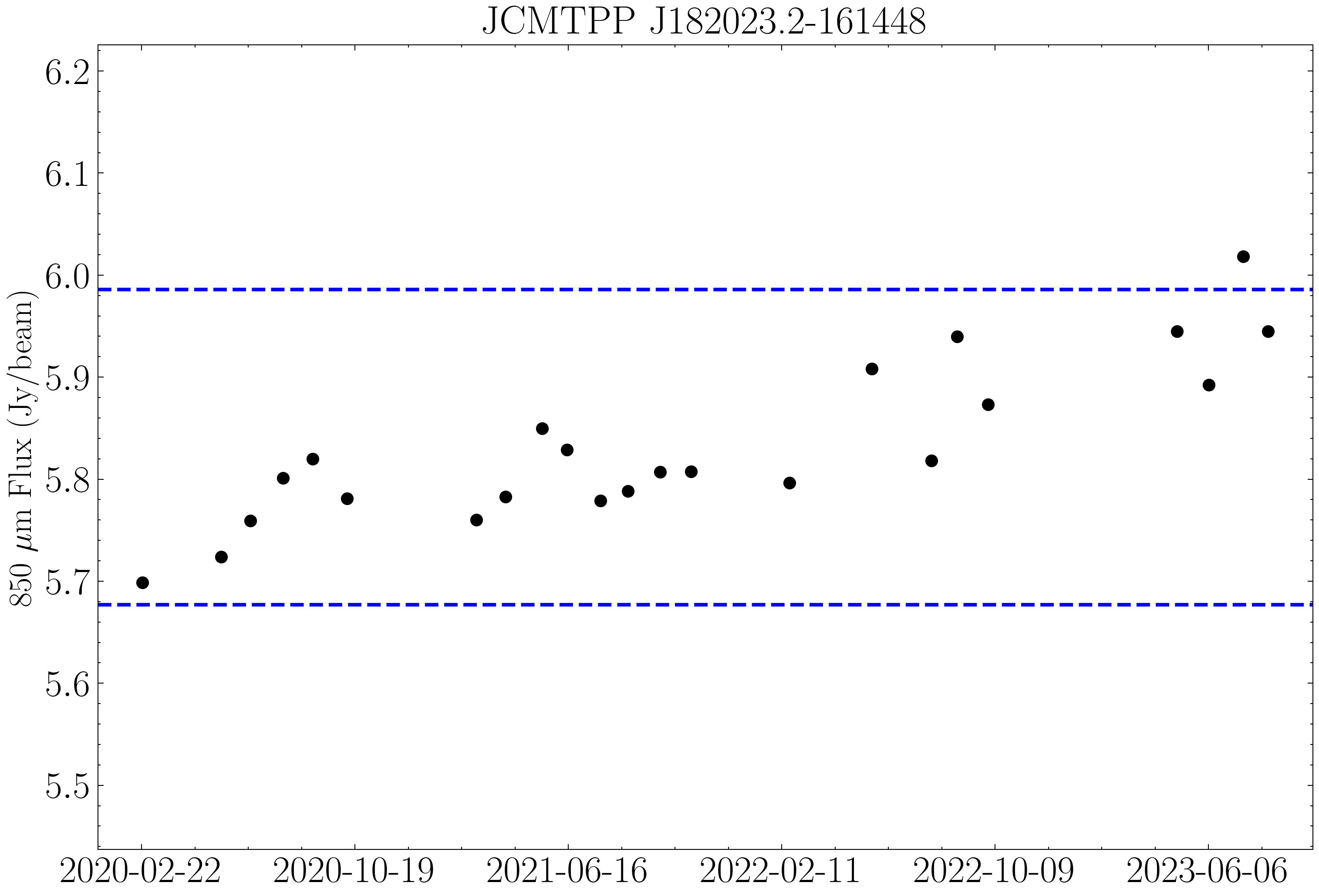}
\caption{$850\,\mu$m light curve of the candidate linear variable JCMTPP\_J182023.2. The blue dashed lines represent the upper and lower boundaries of $\pm2 SD_\mathrm{{fid}}$, where $SD_\mathrm{{fid}}$ is the expected fiducial standard deviation.}
\label{fig:145_lc}
\end{figure}

A bright $850\,\mu$m source, JCMTPP\,J182023.2, with an $|S|/\Delta S=7.8$ stands out as a linear variable in the right panel of Figure~\ref{fig:slope}, with a total linear increase of $\sim 4-5\%$ over 3.5 years. The $850\,\mu$m lightcurve of JCMTPP\,J182023.2 is shown in Figure~\ref{fig:145_lc}. JCMTPP\,J182023.2 is a bright and compact continuum source observable from submillimeter to millimeter wavelengths and associated with a 22-GHz H$_2$O maser \citep{1993MNRAS.264.1025H,2005MNRAS.363..405H,2008ApJS..175..277D,2011MNRAS.416..178B}. 
The $450\,\mu$m counterpart of JCMTPP\,J182023.2 has $|S|/\Delta S=1.27$. Given the relatively high flux calibration uncertainty at $450\,\mu$m of $\sim3$\% derived earlier, an observed fractional change of $\sim1$\%/yr would not be recovered before many years of observation.

\begin{figure}
\centering
\includegraphics[width=0.45\textwidth]{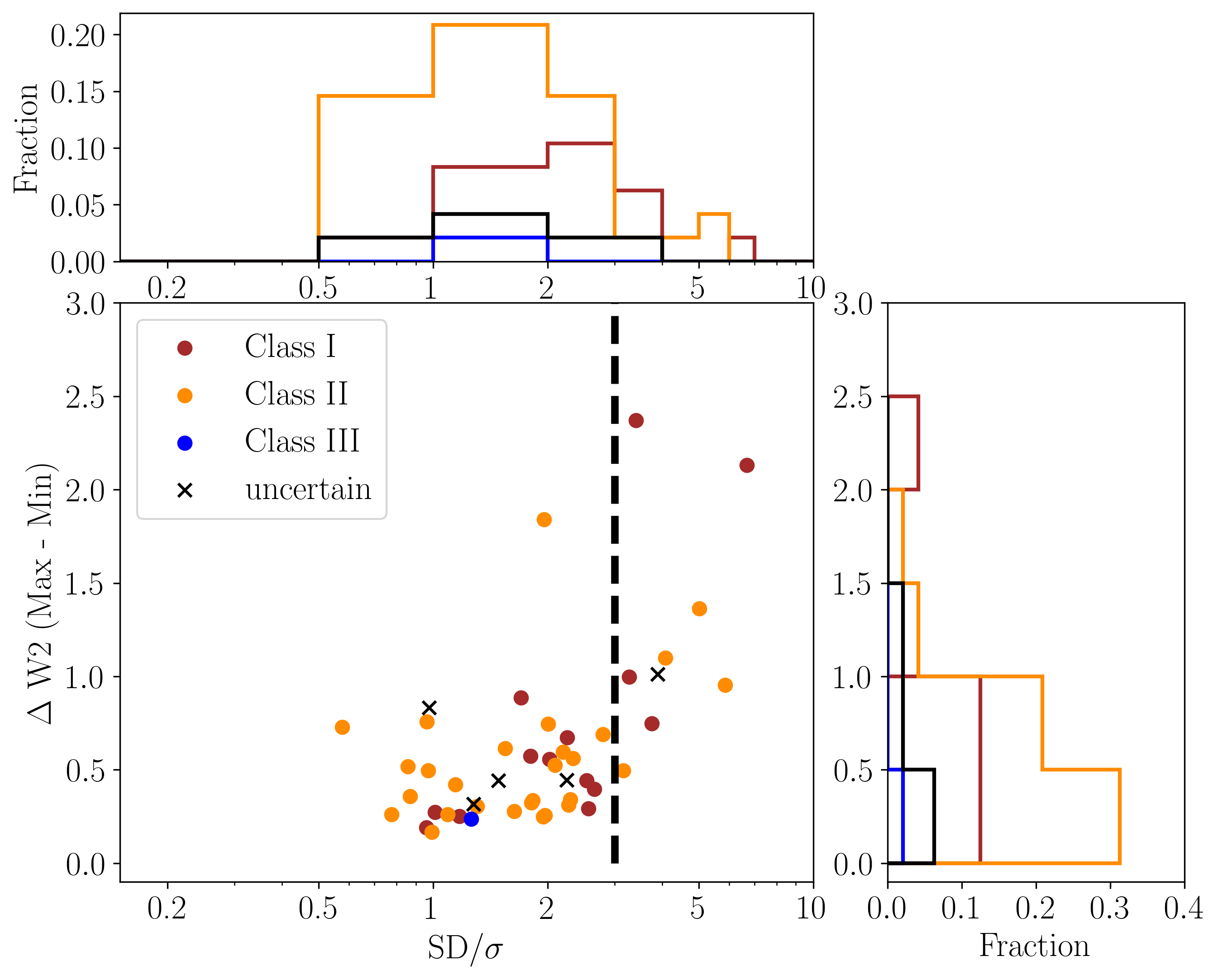}
\caption{The distribution of the $\Delta W2$ versus $SD/\sigma$ for the YSO sample with $\Delta \mathrm{W2}/\sigma \mathrm{(W2)} > 3$. The vertical dashed line indicates the commonly used threshold value of $SD/\sigma=3$ for identifying variable sources.}
\label{fig:MIR_SD}
\end{figure}

\subsection{Searching for variable YSOs in the mid-IR}\label{sect:mir}
We uncovered a sample of 66 YSOs in M\,17 with reliable multi-epoch data in the \textit{WISE} W1 and/or W2 bands. Some of the deeply embedded YSOs, for example M17\,MIR \citep{2021ApJ...922...90C}, are very faint or not detected at all in the W1 band. Therefore, and following \citet{2021ApJ...920..132P}, the mid-IR light curve analysis for the majority of the sample is based on the multi-epoch data in the W2 band, supplemented with analysis of W1 data for specific sources. 

For the 66 YSOs with at least 12 epochs of photometry, we obtained their variability amplitude of W2 magnitude $\Delta \mathrm{W2}$, the difference between the maximum and minimum magnitudes, and their mean uncertainty of W2 magnitude $\sigma \mathrm{(W2)}$. \citet{2021ApJ...920..132P} employed a criterion of $\Delta \mathrm{W2}/\sigma \mathrm{(W2)} > 3$ to exclude sources that are unlikely to be variable and to search for variability in the sources remaining. After applying this criterion, they divided the mid-IR variability of YSOs in the Gould Belt into two major types, secular and stochastic variables. According to their mid-IR light curves, the secular variables can be further split into three categories, linear, periodic, and curved; while the stochastic variability includes burst, drop, and irregular \citep[see definitions in][]{2021ApJ...920..132P}.  

We first examine YSOs with $\Delta \mathrm{W2}/\sigma \mathrm{(W2)} > 3$ in the magnitude domain to find variability. Within the sample of 66 YSOs with NEOWISE data in at least 12 epochs, 48 sources satisfy this criterion. We applied the same methods as in \citet{2021ApJ...920..132P}, to these 48 YSOs to investigate their types of variability. 

We estimate a flux standard deviation ($\mathrm{SD}$) and a mean flux uncertainty ($\sigma$) for the 48 YSOs with   $\Delta \mathrm{W2}/\sigma \mathrm{(W2)} > 3$. Figure~\ref{fig:MIR_SD} the distribution of these YSOs in the diagram of $\Delta \mathrm{W2}$ versus $\mathrm{SD}/\sigma$. Similarly as \citet{2021ApJ...920..132P}, we employed the criterion of $\mathrm{SD}/\sigma >3$ to classify variables in general. Nine YSOs have $\mathrm{SD}/\sigma > 3$, and are all at younger stages (Class I, II), with the exception for one source (SPICY\,82001) which has an uncertain classification.\footnote{This source was classified as uncertain  by \citet{2009ApJ...696.1278P,2021ApJS..254...33K}, while \citet{2013ApJS..209...31P} classified it as a Class II/III YSO.}  

Some YSOs in Figure~\ref{fig:MIR_SD} show large brightness variations despite having $\mathrm{SD}/\sigma$ values less than 3. Previously, \citet{2021ApJ...920..132P} found that some types of YSO variability may not lead to a large $\mathrm{SD}$ over the full light curve. In order to classify as many variables as possible, we also applied the Lomb-Scargle periodogram (LSP) \citep{1976Ap&SS..39..447L,1989ApJ...343..874S} to the mid-IR light curves using the \textit{LombScargle} from \textit{python} package \textit{astropy}. Since the WISE/NEOWISE surveys have a 6 month cadence, periodic variations with periods shorter than 6 months cannot be extracted from the data. Furthermore, light curves with periods longer than 2300 days cover fewer than two full phases, since the total duration of the NEOWISE data is about 4600 days (12.5 years). Given this, we cannot conclude that these are variable YSOs with long periodic (periods longer than 2300 days) variability. Periods longer than 2300 days manifest as an increasing or decreasing trend in mid-IR brightness. To quantify the significance of the amplitude and period, we compute the false alarm probability (FAP) of LSP analysis, $\mathrm{FAP_{LSP}}$. This provides the uncertainty of a particular LSP peak by quantifying the probability of a false peak due to random errors. We slightly modified the method by \citet{2008MNRAS.385.1279B} to determine the FAP of the found period or longer, rather than summing over all periods within the range checked \citep[see also][]{2021ApJ...920..132P,2021ApJ...920..119L}. 

We further adopted a linear least-squares fit to find a linear trend of increasing or decreasing fluxes. Light curves with good linear fits are also often fitted by LSP with a very long period. We define the linear FAP (hereafter $\mathrm{FAP_{Lin}}$, with the same method as \citet{2021ApJ...920..119L}, to estimate the likelihood of the determined best-fit linear slope. If the value of $\mathrm{FAP_{Lin}}$ of a given source is $<10^{-4}$, the light curve of this source can be robustly fitted by a linear slope. For LSP analysis, a somewhat lower threshold, $\mathrm{FAP_{LSP}}<10^{-2}$, is used to explore the wide range of periods and amplitudes recovered. This lower threshold results in a few false positives within the LSP sample; however, we have checked to ensure that these false positives result only in a small contamination fraction. 

\begin{figure}
\centering
\includegraphics[width=0.45\textwidth]{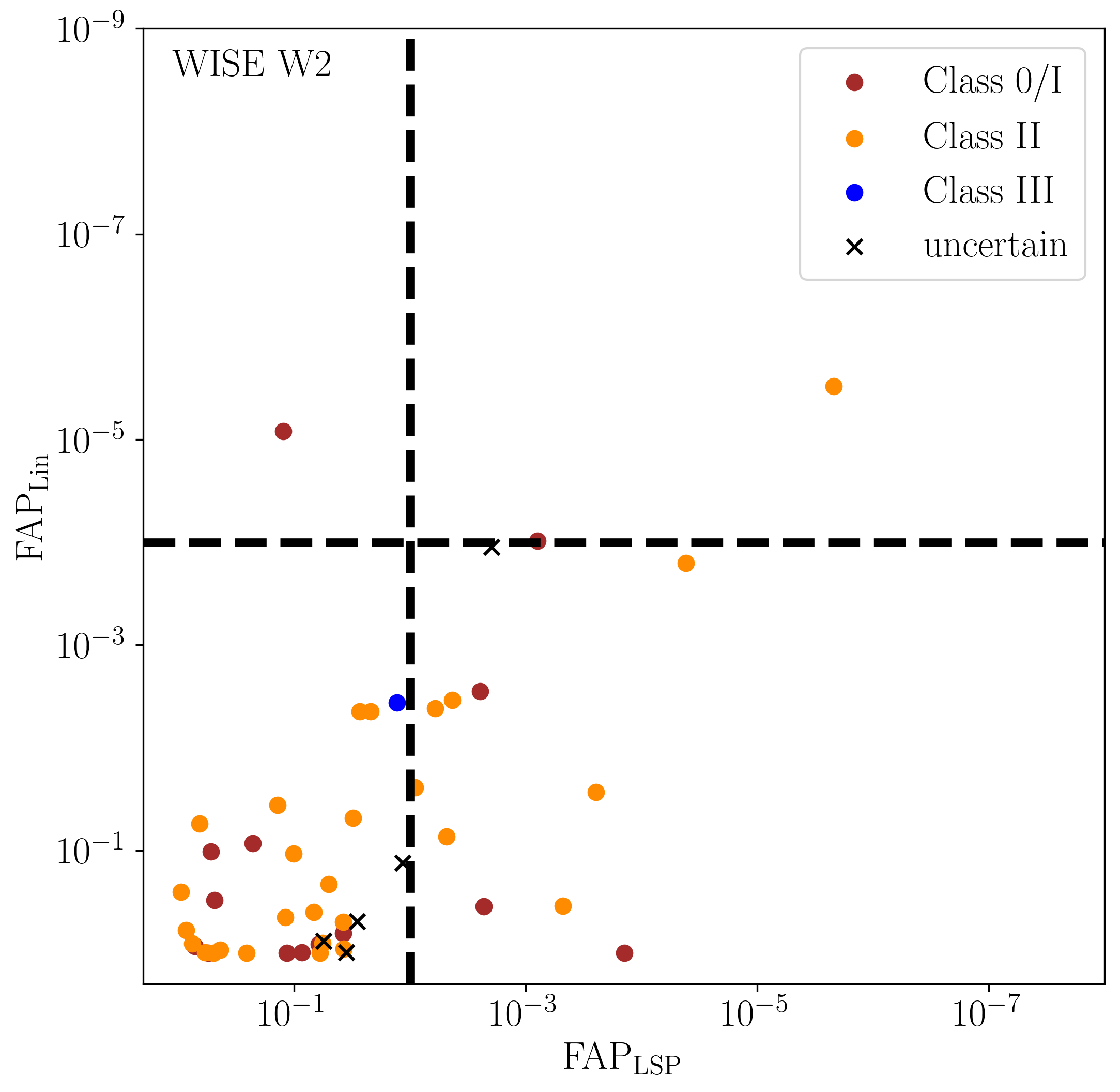}
\caption{Comparison of the FAPs for linear and periodic fits for the same YSO as in Figure~\ref{fig:MIR_SD}. The vertical dashed line represents a FAP value of $10^{-4}$ for linear fit, while the horizontal dashed line represents a FAP value of $10^{-2}$ for LSP fit. }
\label{fig:MIR_Linear}
\end{figure}

In Figure~\ref{fig:MIR_Linear}, we consider both the periodic and linear FAPs for the best fits to the 48 YSOs with $\Delta \mathrm{W2}/\sigma \mathrm{(W2)} > 3$. Sources with $\mathrm{FAP_{Lin}} \leqslant 10^{-4}$ lying in the upper region in Figure~\ref{fig:MIR_Linear} are classified as showing linear variability. For the remaining sources with $\mathrm{FAP_{LSP}} \leqslant 10^{-2}$, those with a period of 2300 days or less are classified as periodic variables, and those with a longer period are classified as showing curved variability. Finally, 14 out of the initial 48 candidate variables are classified as curved (6), periodic (6) and linear (2). In general, we categorize them as secular variables.

The candidate variable sources that fail the conditions of secular variability are classified as stochastic. An additional constraint is applied to identify sources that exhibit bursts or drops in brightness at some epochs while maintaining stable fluxes over the remaining epochs. Sources with median(W2) - min(W2) $> 0.8\times \Delta \mathrm{W2(max-min)}$ are burst type, and those with max(W2) - median(W2) $> 0.8\times \Delta \mathrm{W2(max-min)}$ are drop type. After excluding all previously identified sources, those with $\mathrm{SD}/\sigma > 3$ are classified as irregular variables. In particular, we reclassify the outbursting MYSO M17~MIR as irregular due to a high value of $\mathrm{SD}/\sigma=6.69$, despite the roughly linear trend ($\mathrm{FAP_{Lin}}=8.28e-06$) of its mid-IR light curve (see also Figure~\ref{fig:lc_M17MIR}). 

Following the taxonomy of mid-IR variability \citep{2021ApJ...920..132P}, we classify the variability types for 22 out of the 48 YSOs with $\Delta \mathrm{W2}/\sigma \mathrm{(W2)} > 3$. The remaining 26 YSOs cannot pass any criterion for secular and stochastic variables. Table~\ref{tab:variables_mir_type} summarizes the statistical results of the types of variability along with the evolutionary stages for the 22 YSOs with mid-IR variability. For completeness, the statistics of all 22 variable YSOs are presented in Table~\ref{tab:mir_variable}. Two YSOs (\#6 and \#10 in Table~\ref{tab:mir_variable}) show more pronounced variability in W1 than in W2, so their variability types are classified from the W1 band. 

Fourteen of the 22 variable YSOs in M\,17 identified in this study are secular variables, while the remaining 8 stochastic variables are dominated by irregular types. Almost all (19) of the 22 variables are Class I and II YSOs, only three variable YSOs are of uncertain stage. Although the number of Class I variables is smaller than that of Class II variables, the fraction (38.1\%) of Class I YSOs with mid-IR variability is higher than that (29.7\%) of the Class II YSO sample.

\begin{deluxetable*}{cccccc}
\tablecaption{MIR Variability Type by YSO Classification \label{tab:variables_mir_type}}
\tablehead{
\colhead{Variability type} & \colhead{Class I} & \colhead{Class II} & \colhead{Class III} & \colhead{uncertain} & \colhead{Total}  }
\startdata
    Linear & 1  (4.8) & 1  (2.7) & 0   & 0   & 2 \\
    Curved & 2 (9.5) & 3  (8.1) & 0   & 1  (14.3) & 6 \\
    Periodic & 2  (9.5) & 3  (8.1) & 0 & 1 (14.3)  & 6 \\
    Burst & 1  (4.8) & 2  (5.4) & 0   & 0  & 3 \\
    Drop & 0 & 0   & 0   & 0   &  0 \\
    Irregular & 2  (9.5)& 2  (5.4) & 0   & 1  (14.3) & 5 \\
    \hline
    Total & 8 (38.1) & 11  (29.7) & 0  & 3  (42.9) & 22 \\
    \hline
    All YSOs & 21 & 37 & 1 & 8  & 66\\
\enddata
\tablecomments{Numbers are the count of variables for each variable type, while numbers in parentheses are the fraction (\%) of variable candidates relative to the selected samples in each evolutionary stage .}
\end{deluxetable*}

\begin{figure*}
\centering
\includegraphics[width=0.9\textwidth]{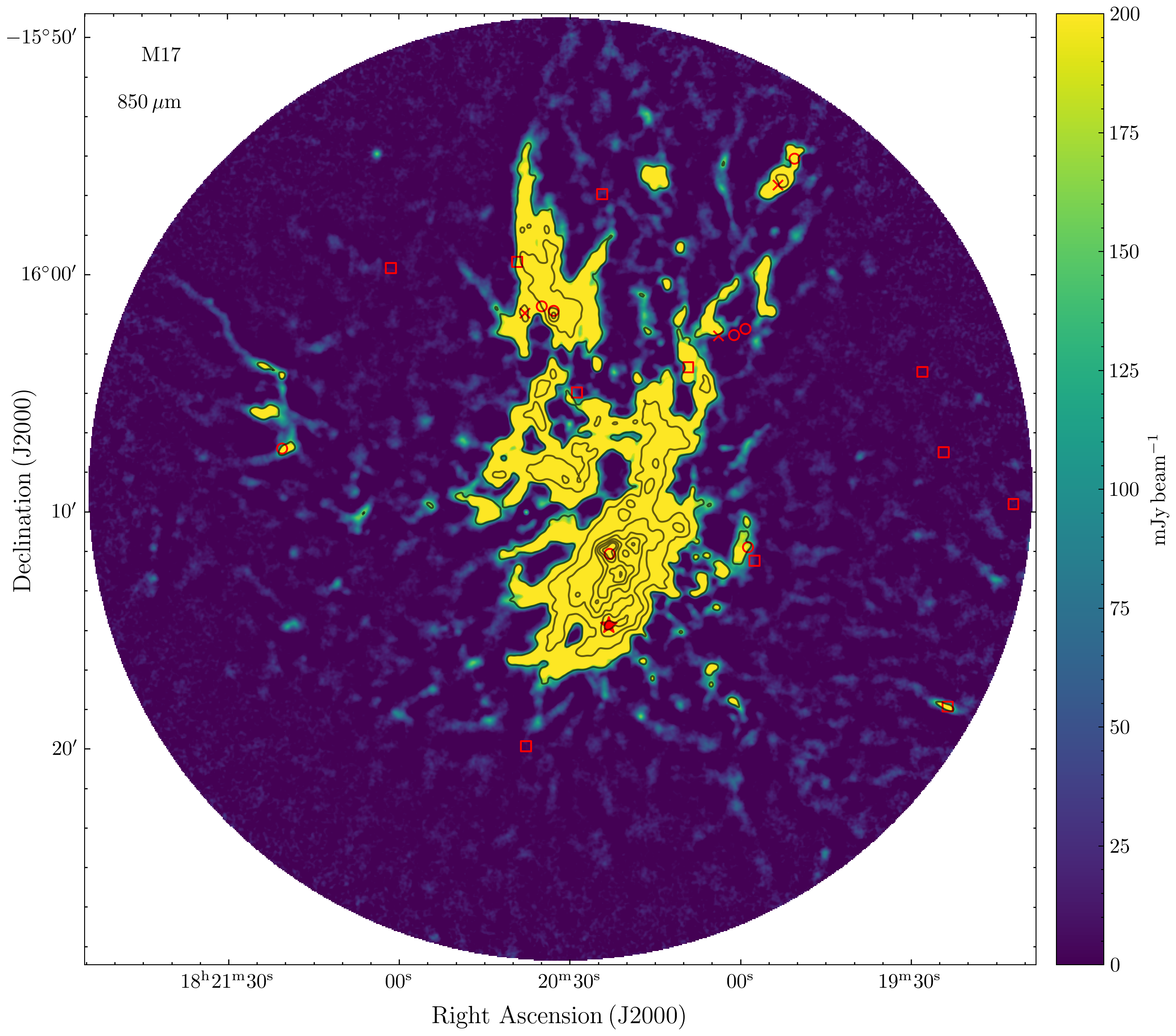}
\caption{The spatial distribution of mid-IR and submillimeter variable candidates classified in this work, overlaid on the $850\,\mu$m map of M\,17. The contours of $850\,\mu$m continuum emission are in levels of 0.15, 0.6, 1.8, 3.6, 6.0, 9.0, 1.26 and $1.68\,\mathrm{Jy\,beam^{-1}}$. Figure~\ref{fig:coadd}. Circle, square, and `X' represent different classification type for YSOs, i.e., Class I, Class II, and uncertain, respectively. The star symbol represents the linear variable at $850\,\mu$m (JCMTPP\_J182023.2). }
\label{fig:850_MIR_sources}
\end{figure*}

\begin{figure*}
\centering
\includegraphics[width=0.8\textwidth]{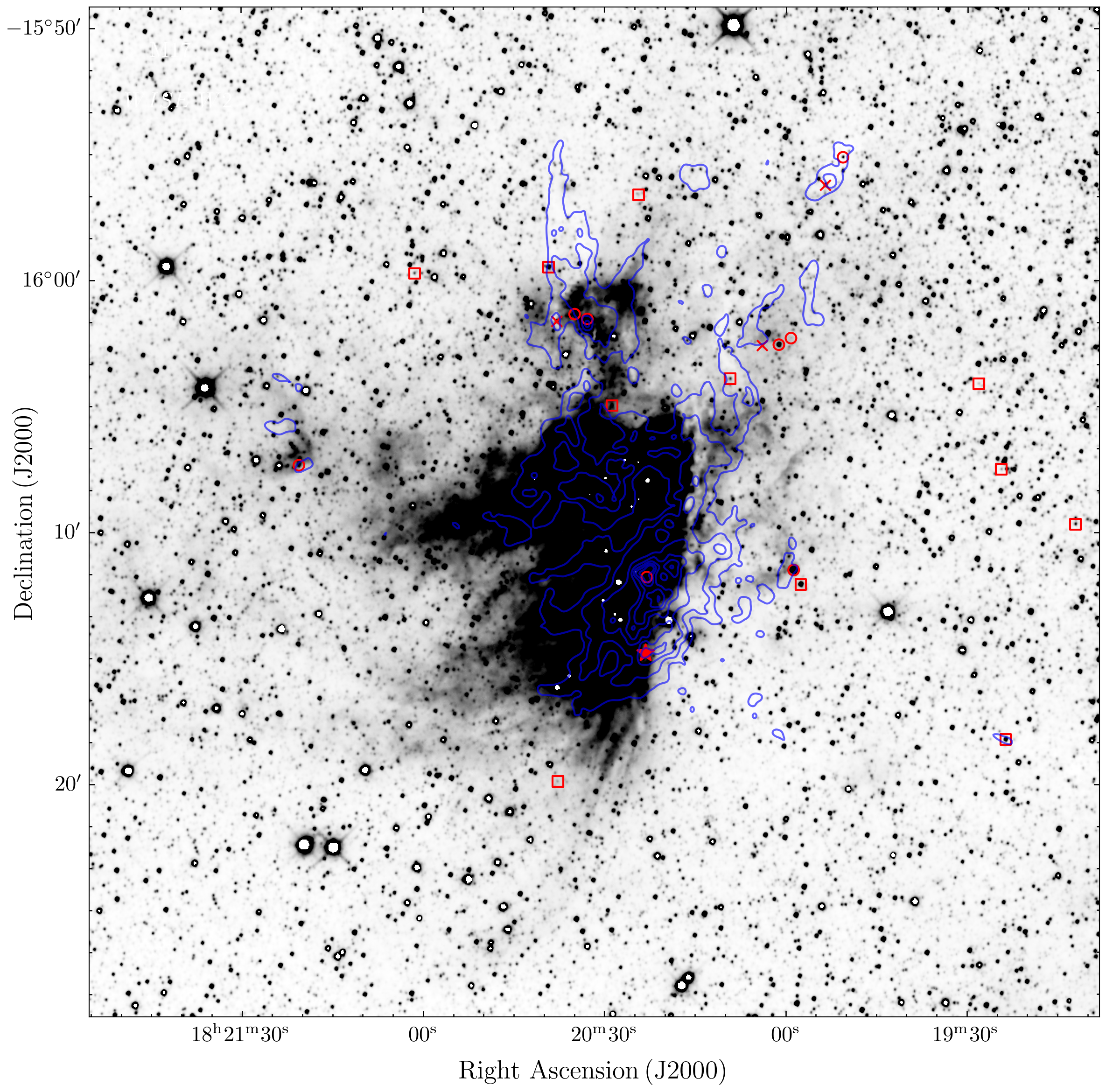}
\caption{Same as Fig.~\ref{fig:850_MIR_sources}, but for the mid-IR \textit{WISE} W2 map in gray scale. The contours of $850\,\mu$m continuum emission are as same as in Figure~\ref{fig:850_MIR_sources} but are displayed as blue.}
\end{figure*}

\subsection{Cross-matching the submillimeter sources with the variable YSOs}\label{sect:submm+MIR}
For the 22 YSOs showing mid-IR variability, we searched for the nearest submillimeter sources at 450 and $850\,\mu$m.
Four variable YSOs have bright counterparts at $450\,\mu$m and six variable YSOs have bright counterparts at $850\,\mu$m within a radius of $10\arcsec$ (Table~\ref{tab:mir_submm_sources}), following the distance constraint applied by \citealt{2020MNRAS.495.3614C}. 
We manually checked the spatial distribution of the 22 variable YSOs overlaid on the 450 and $850\,\mu$m maps. Expanding the search radius to $15\arcsec$ leads to one additional potential submillimeter counterpart for SPICY\,81957.


Among the seven variable candidate YSOs with submillimeter counterparts, four YSOs (SPICY\,81351, 82315, 81642, and 81623) are located on the outskirts of the M\,17 region. Two of these candidate YSOs, SPICY\,81642 and 81623, have periods of 437.8 and $>4800$ days, respectively. Both 
have an uncertain class for their YSO classification \citep{2013ApJS..209...31P,2021ApJS..254...33K}. We suggest SPICY\,81642 and 81623 are likely contamination from evolved post-main-sequence stars with periods typically longer than hundreds of days \citep[e.g.,][]{2021AJ....162...52Y}. The remaining two variables (SPICY\,81351 and 82315) have an irregular variability type and are closer to their nearby submillimeter peaks than the former two. We suggest that SPICY\,81351 and 82315, as well as their submillimeter counterparts, are parts of the M\,17 complex, but spatially distinct from the M\,17 \ion{H}{2} region. The remaining three variable YSO candidates, SPICY\,81957 and 82001, and M17~MIR, are located at the two star-forming clouds M17 North and SW, respectively. We will discuss them in more detail in Sect.~\ref{sect:mir_variables}.


\begin{table*}[]
    \centering
    \caption{Matched variable YSO candidates and $450$ or $850\,\mu$m sources within a radius of $15\arcsec$.}
    \begin{tabular}{c c c c c c}
    
    \hline\hline
   \# &    Variable YSO   &  $450\,\mu$m source & Separation ($\arcsec$) & $850\,\mu$m source & Separation ($\arcsec$) \\
      \hline
    2 &  SPICY\,81351  & JCMTPP\_J181923.8 & 3.1 & JCMTPP\_J181923.4 & 5.2 \\
    5  &  SPICY\,81623  & -                 &   -   & JCMTPP\_J181950.6 & 7.0 \\
    6   & SPICY\,81642  & JCMTPP\_J181954.0 & 12.0 & JCMTPP\_J181953.9 & 8.4 \\
    15 & SPICY\,81957  & JCMTPP\_J182032.9 & 15.0 & JCMTPP\_J182032.8 & 12.0 \\
    18  & SPICY\,82001  & JCMTPP\_J182038.1 & 6.2 & JCMTPP\_J182038.2 & 5.6 \\
    21  & SPICY\,82315  & JCMTPP\_J182120.2 & 6.4 & JCMTPP\_J182120.1 & 6.9 \\
    22  & M17~MIR       & JCMTPP\_J182022.7 & 9.2 & JCMTPP\_J182022.8 & 7.0 \\
        \hline  
    \end{tabular}
    
    \label{tab:mir_submm_sources}
\end{table*}

 


\begin{figure}
\centering
\includegraphics[width=0.45\textwidth]{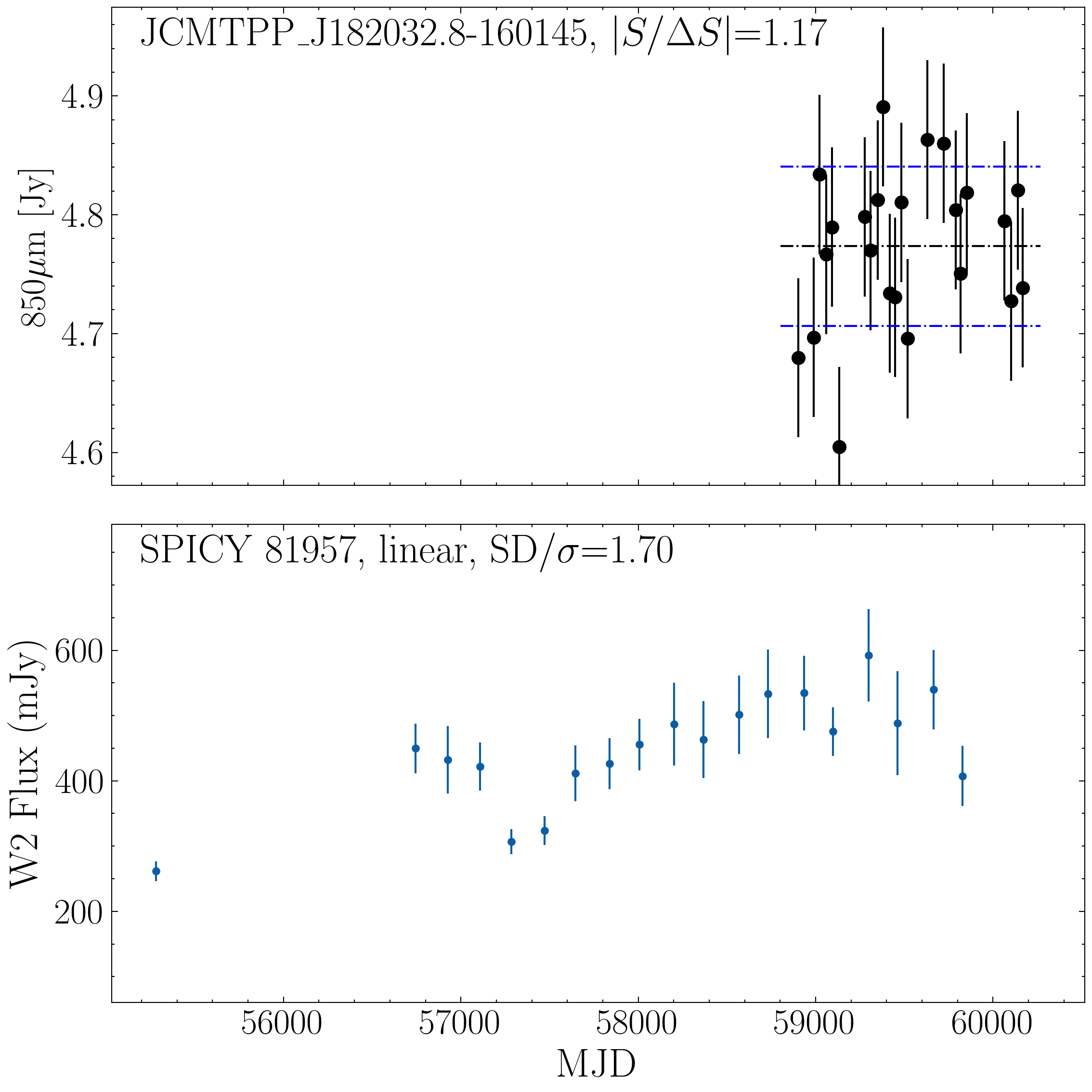}
\caption{JCMT Transient $850\,\mu$m (top) and NEOWISE W2 light curves (bottom) of SPICY\,81957. The black and blue dashed lines in the JCMT Transient light curve represent the mean peak flux and $\pm1\sigma$ boundaries.}
\label{fig:lc_81957}
\end{figure}

\begin{figure}
\centering
\includegraphics[width=0.45\textwidth]{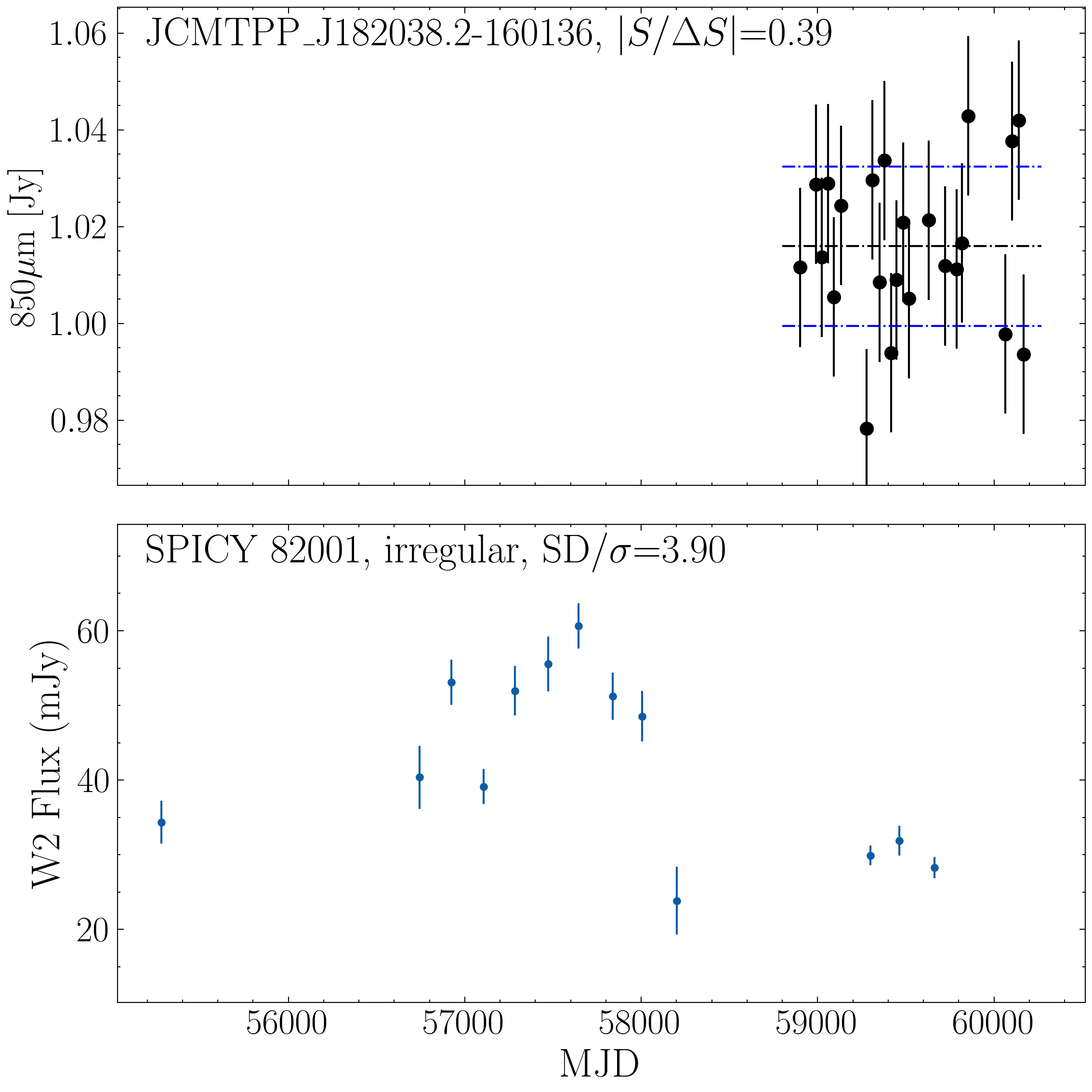}
\caption{Same as Figure~\ref{fig:lc_81957} but for SPICY\,82001.}
\label{fig:lc_82001}
\end{figure}

\begin{figure}
\centering
\includegraphics[width=0.45\textwidth]{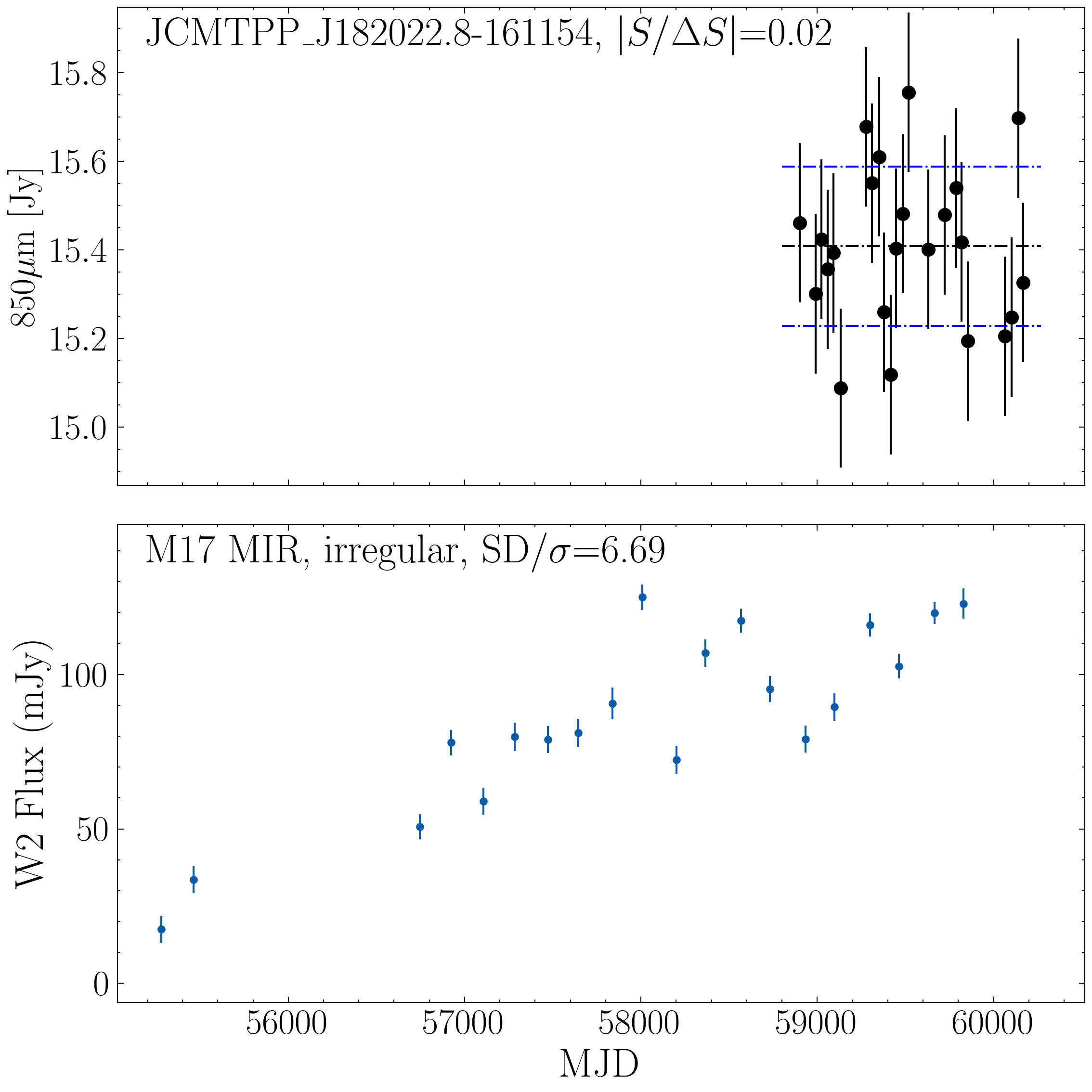}
\caption{Same as Figure~\ref{fig:lc_81957} but for M17~MIR.}
\label{fig:lc_M17MIR}
\end{figure}

\begin{figure*}
\centering
\includegraphics[width=0.9\textwidth]{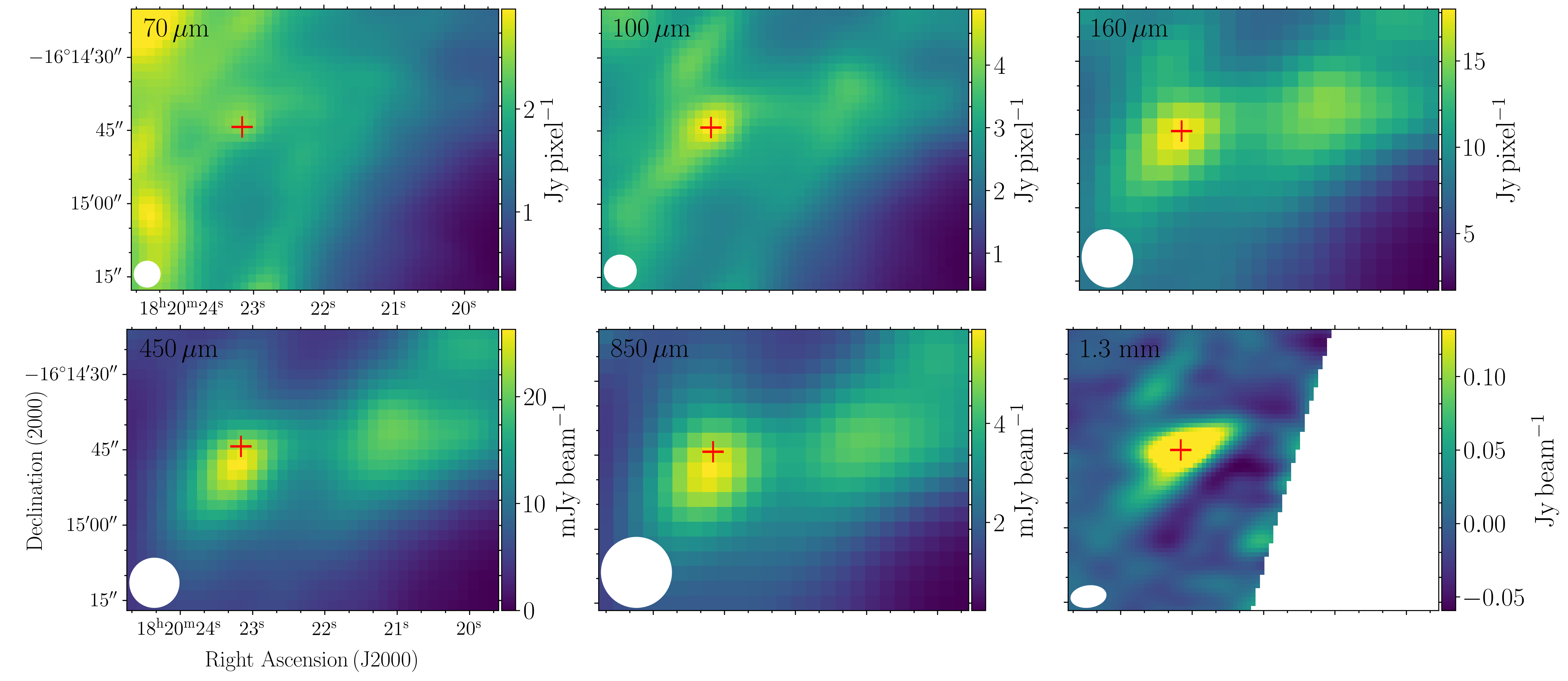}
\caption{Multiwavelength maps for JCMTPP\_J182023.2, collected from \textit{Herschel} (70, 100, and $160\,\mu$m), JCMT Transient Survey (450 and $850\,\mu$m), and 1.3 mm continuum by ACA. The red cross denotes the position of 22 GHz H$_2$O maser source \citep{2011MNRAS.416..178B}.}
\label{fig:145_multi}
\end{figure*}




\begin{figure}
\centering
\includegraphics[width=0.45\textwidth]{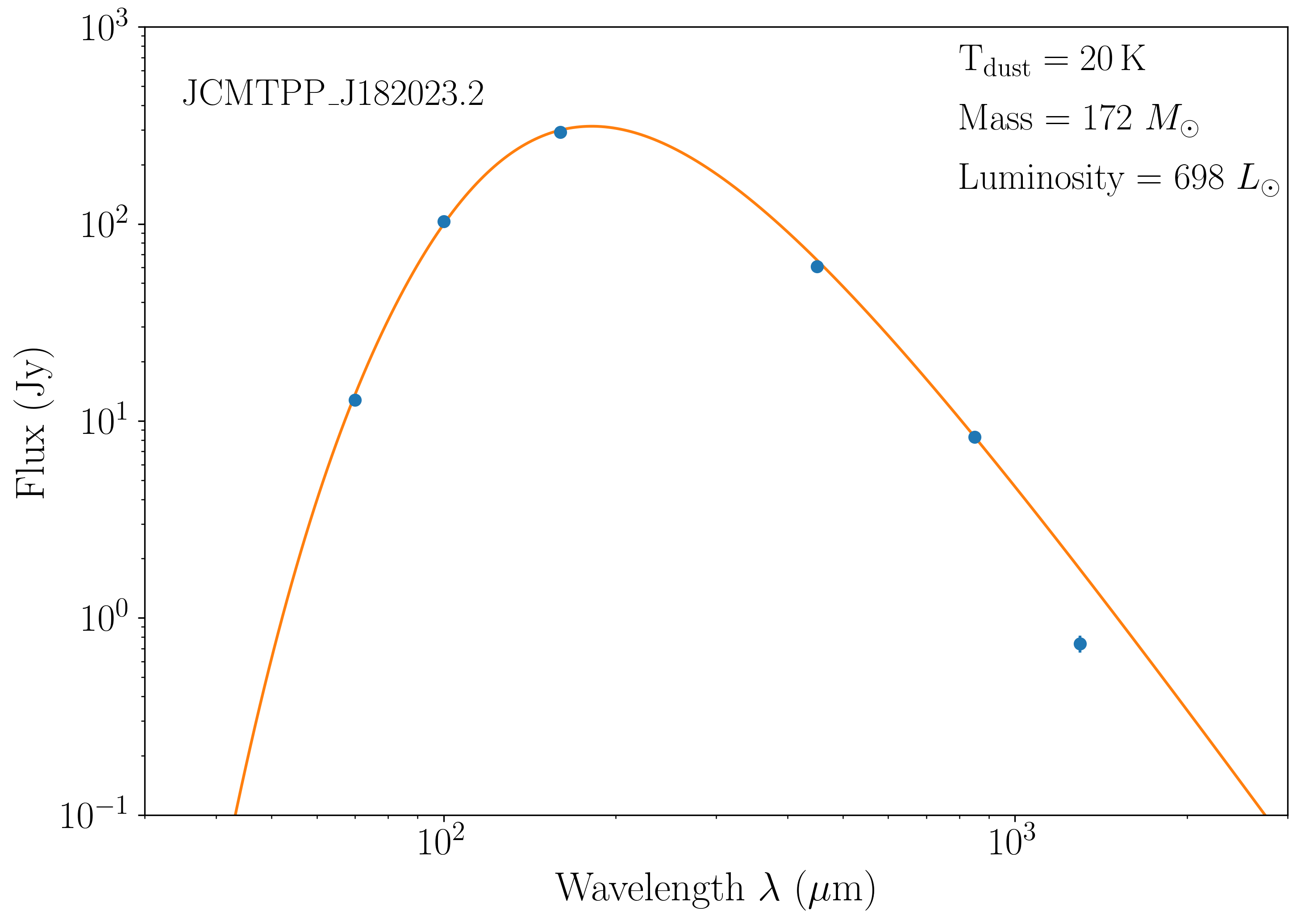}
\caption{Spectral energy distribution of the linear variable JCMTPP\_J182023.2. The best-fit greybody model is shown as the orange curve.}
\label{fig:145_SED}
\end{figure}

\section{Discussion}\label{sect:discussion}
\subsection{Interesting YSOs with mid-IR variability}\label{sect:mir_variables}
SPICY\,81957 is located at the boundary of a bright JCMT source seen at 450 and $850\,\mu$m, the strongest submillimeter peak in M17 North, labeled as M17N by \citet{2006ApJ...644..990R}. H$_2$O maser emission at 22.2 GHz
was detected towards this submillimeter source  \citep{1981ApJ...250..621J}. SPICY\,81957 is the brightest variable YSO in the mid-IR found in this study, modeled as a candidate MYSO of $5-10\,M_\sun$ \citep{2009ApJ...696.1278P}. The JCMT Transient $850\,\mu$m and WISE/NEOWISE W2 light curves of SPICY\,81957 are presented in Figure~\ref{fig:lc_81957}.
SPICY\,81957 shows significant mid-IR variability over $\gtrsim10$ years period. Its submillimeter light curve provided by JCMT Transient Survey covers only the recent $\sim3.5$ years, a small fraction of the NEOWISE timescale. 
The mid-IR flux of SPICY\,81957 is roughly stable throughout most of the period of JCMT Transient survey, except for 2022 September when the W2 flux decreased by a factor of $\sim50\%$ compared to that in 2022 March. However, the presumably linear decrease of $850\,\mu$m peak flux of the associated submillimeter source (JCMTPP\_J182032.8) between 2022 September and 2023 August cannot be confirmed from the JCMT Transient survey data collected in 3.5 years.


SPICY\,82001 is associated with the second brightest submillimeter peak in M17 North. The evolutionary class of SPICY\,82001 is still in debate; \citet{2013ApJS..209...31P} classified it as a Class II/III YSO, while the SPICY catalog assigned an ambiguous type \citep{2021ApJS..254...33K}. Figure~\ref{fig:lc_82001} shows the $850\,\mu$m and WISE/NEOWISE W2 light curves for this source. SPICY\,82001 has been at its faintest mid-IR phase over the period of JCMT Transient survey. The $850\,\mu$m light curve of the associated submillimeter source JCMTPP\_J182038.2 is also flat, in line with expectation.

M17~MIR is associated with the second brightest JCMT source at 450 (JCMTPP\_J182022.7) and $850\,\mu$m (JCMTPP\_J182022.8) throughout the M\,17 field. M17~MIR is located at the M17 SW cloud, adjacent to the M\,17 \ion{H}{2} region. \citet{2021ApJ...922...90C} found that M17~MIR has produced two accretion outbursts in recent decades, one major accretion outburst in the 1990s and one ongoing moderate accretion outburst since mid-2010. \citet{1993MNRAS.264.1025H} derived a total mass of $330\,M_\sun$ and a luminosity of $8000\,L_\sun$ for the natal clump of M17~MIR (FIR3 in their work) using multi-wavelength maps ranging from 450 to $1300\,\mu$m. The $850\,\mu$m and WISE/NEOWISE W2 light curves of M17~MIR are shown in Figure~\ref{fig:lc_M17MIR}. M17~MIR shows significant mid-IR variability over $\gtrsim10$ years period. During the JCMT Transient Survey period, the mid-IR variability of M17~MIR is mild, at a level of $\sim+50\%$. In contrast, the $850\,\mu$m light curve of M17~MIR suggests no submillimeter variability in the same period. 

There are a number of potential reasons for the discrepancy between the mid-IR variability and the currently stable submillimeter peak flux of the three YSOs. First, based on the NEOWISE and JCMT Transient Survey for the Gould Belt, \citet{2020MNRAS.495.3614C} found that 39 YSOs are variable in at least one of the two surveys. However, only 14 of the 39 YSOs show correlated secular variability in mid-IR and submillimeter wavelengths. Second, given the farther distance compared to the Gould Belt, a single JCMT beamsize at $850\,\mu$m encompass more gas and dust for the protostars in M\,17 than for those in the Gould Belt. This is likely to have an impact on the submillimeter variability and will be discussed below in Sect.~\ref{sect:sub_variable} alongside the only secular submillimeter variable discovered in M\,17 and in Sect.~\ref{sect:comparison} compared to other fields of JCMT Transient Survey.

\subsection{Candidate linear variable at $850\,\mu$m in M\,17}\label{sect:sub_variable}
JCMTPP\_J182023.2 is the only candidate linear variable at $850\,\mu$m recovered by the JCMT Transient Survey for M\,17. 
This source, however, is not visible in the mid-IR with either \textit{Spitzer} or \textit{WISE}. The protostar(s) embedded within JCMTPP\_J182023.2 is therefore still infrared faint, and the soure is likely at an earliest stage.

Figure~\ref{fig:145_multi} presents continuum maps centered on JCMTPP\_J182023.2 at far-IR through millimeter wavelengths, including \textit{Herschel} 70, 100, $160\,\mu$m, JCMT-SCUBA2 450 and $850\,\mu$m, and ACA 1.3 mm. JCMTPP\_J182023.2 starts to appear in the \textit{Herschel} $70\,\mu$m image, although it remains still faint. Interestingly, the 22 GHz H$_2$O maser emission \citep{2011MNRAS.416..178B}, denoted by the red cross in all maps, spatially coincides with JCMTPP\_J182023.2. Very young and embedded protostars are expected to power bipolar jets, which physically interact with ambient materials in their immediate environment \citep{2024ApJ...969L...6Z}. In particular, 22.2\,GHz H$_2$O maser emission emerges from the shocked molecular gas at the interface between the fast flow and the ambient material, thus efficiently tracing protostellar outflows \citep[e.g.,][]{2005A&A...438..571F,2005ApJS..156..179D,2013A&A...549A.122M}. The 22.2 GHz H$_2$O maser emission associated with JCMTPP\_J182023.2 suggests that this compact submillimeter source might host one or more deeply embedded protostar(s) that power protostellar powerful outflow(s). We also note a signature of molecular outflow for the CO ($2-1$) line emission around JCMTPP\_J182023.2 from the ACA band 6 data. The integrated map of the CO ($2-1$) line wing emission clearly show a bipolar molecular outflow roughly perpendicular to the orientation of the elongated core seen in 1.3 mm dust continuum. 

The integrated flux of JCMTPP\_J182023.2 at multiple wavelengths is extracted using an aperture size equal to the beam size of each map. Figure~\ref{fig:145_SED} shows the spectral energy distribution (SED) of JCMTPP\_J182023.2 constructed from the multiwavelength fluxes. We also obtain a best-fitting greybody  dust temperature of $20\,\mathrm{K}$ from the SED fitting procedure `CMCIRSED' developed by \citet{2012MNRAS.425.3094C}.



Assuming the submillimeter and millimeter emission is optically thin, the observed submillimeter and millimeter emission of JCMTPP\_J182023.2 can be used to derive the mass of this source using
\begin{equation}
M=\frac{S^\mathrm{int}_\lambda\,d^2}{\kappa_\lambda\,B_\lambda(T_\mathrm{d})},  
\end{equation}
where $S^\mathrm{int}_\lambda$ is the integrated flux at wavelength $\lambda$, $d$ is the distance to M\,17, $\kappa_\lambda$ is the dust opacity per unit mass at wavelength $\lambda$, and $B_\lambda(T_\mathrm{d})$ is the Planck function evaluated at the dust temperature $T_\mathrm{d})$. For this work, we choose $d=2.0\,\mathrm{kpc}$ \citep{2011ApJ...733...25X}, $\kappa_{850}=0.019\,\mathrm{cm^2\,g^{-1}}$ at $850\,\mu$m and $\kappa_{1.3}=0.013\,\mathrm{cm^2\,g^{-1}}$ at 1.3 mm \citep{2021MNRAS.501.1316L}, and $T_\mathrm{d}=20\,\mathrm{K}$. The values of $\kappa_{850}$ and $\kappa_{1.3}$ are converted to opacity relative to gas mass by assuming a standard gas-to-dust ratio of 100. The integrated flux of JCMTPP\_J182023.2 is $8.3\,\mathrm{Jy}$ and $0.74\,\mathrm{Jy}$ at $850\,\mu$m and 1.3 mm, respectively. This source is also detected at 1.2-mm continuum emission with an FWHM of $43\arcsec$ by SIMBA observations with the Swedish ESO Submillimeter Telescope \citep{2005MNRAS.363..405H} and the authors report an integrated flux of 3.6 Jy and a peak flux of $1.6\,\mathrm{Jy\,beam^{-1}}$.
Thus, we warn that the 1.3 mm ACA observations might recover only a small fraction of the total flux. The mass of JCMTPP\_J182023.2 derived from its $850\,\mu$m integrated flux is $172\,M_\sun$, which characterizes the envelope.  A slightly higher value ($\sim 190\,M_\sun$) for the envelope mass is obtained
from the 1.2 mm SIMBA observations, where $\kappa_{1.2}=0.014\,\mathrm{cm^2\,g^{-1}}$ scaled from $\kappa \propto \lambda ^{-1}$ and $\kappa_{1.3}=0.013\,\mathrm{cm^2\,g^{-1}}$ \citep{2021MNRAS.501.1316L}. 
Meanwhile the core mass of JCMTPP\_J182023.2, traced by the compact source seen at the ACA 1.3 mm continuum emission, is $\sim45\,M_\sun$, roughly a quarter of the envelope mass. 


The bolometric luminosity of JCMTPP\_J182023.2 is $698\,L_\sun$, calculated by integrating the best-fit greybody SED in the range from mid-IR to millimeter and assuming that the radiation from the central protostar is fully absorbed and then redistributed by dust through mid-IR to millimeter wavelengths.
The envelope mass and bolometric luminosity locate JCMTPP\_J182023.2 in the region for the accretion phase characterized by the roughly vertical evolutionary track in the luminosity versus mass diagram \citep{2008A&A...481..345M}. According to the classification of evolutionary stages in massive star formation \citep{2022MNRAS.510.3389U}, JCMTPP\_J182023.2 is of the protostellar type. The luminosity-to-mass ($L_\mathrm{bol}/M_\mathrm{env}$) ratio of $4\,L_\sun\,M_\sun^{-1}$ for JCMTPP\_J182023.2 is in line with the  $L_\mathrm{bol}/M_\mathrm{env}$-ratio of $\sim0.2-5\,L_\sun\,M_\sun^{-1}$ for the protostellar type and overlaps with the YSO type in the range $\sim0.7-14\,L_\sun\,M_\sun^{-1}$ \citep{2022MNRAS.510.3389U}. The $L_\mathrm{bol}/M_\mathrm{env}$-ratio increases with the evolutionary stage from quiescent to \ion{H}{2} region, and the rate of change of $L_\mathrm{bol}/M_\mathrm{env}$-ratio continues to accelerate as the embedded object evolves through the protostellar and YSO stages, considering statistical analyses \citep{2022MNRAS.510.3389U}. This acceleration of the $L_\mathrm{bol}/M_\mathrm{env}$-ratio with evolution is suggested to be related to an increase in the accretion rate over time for the earliest stages of massive star formation \citep{2022MNRAS.510.3389U}. For this individual source, the secular trend of increasing brightness at $850\,\mu$m over 3.5 yrs of JCMTPP\_J182023.2 implies a mild increase in the accretion rate over yearly timescales. Be aware that such a mild increase of accretion rate estimated from $850\,\mu$m variability is probably a lower limit. Higher resolution observations with submillimeter interferometric arrays (e.g. ALMA, ACA, SMA) for the JCMT Transient variables found in the Gould Belt revealed a few times larger flux variation than the JCMT Transient $850\,\mu$m variability \citep{2022ApJ...937...29F,2025ApJ...982..176S}. Continued submillimeter observations for JCMTPP\_J182023.2 at a higher resolution (e.g. ACA) are essential to confirm this interpretation and determine the connection to the broader evolutionary conclusions made by \citet{2022MNRAS.510.3389U}.



\subsection{Submillimeter flux change as the probe of accretion luminosity outburst}\label{sect:lumin}
Radiative transfer modeling of eruptive YSOs with a broad range of outburst magnitudes show that the SED variation for different outburst luminosities is wavelength dependent. The flux change is directly proportional to the outburst luminosity at around $100\,\mu$m, while the flux change at submillimeter wavelengths is small \citep{2019MNRAS.487.4465M,2024AJ....167...82F}.
As long as the dust temperature remains above $\sim 25\,\mathrm{K}$, above which the Rayleigh-Jeans relation holds for $850\,\mu$m, the submillimeter response will be approximately linear to the envelope temperature variation \citep{2013ApJ...765..133J,2017ApJ...849..107M}. However, at a lower dust temperature of 20 K, the $850\,\mu$m flux varies as $\propto T_d^{1.5}$, a somewhat stronger than linear response due to the fact that at such low temperature the emission at $850\,\mu$m deviates from the Rayleigh-Jeans tail \citep{2020MNRAS.495.3614C}.

The observed difference in the submillimeter flux of embedded protostar(s) undergoing a change in accretion rate is determined by the heating or cooling of the dusty envelope. \citet{2020MNRAS.495.3614C} explored the submillimeter and dust temperature response to accretion luminosity outburst in details, and predicted that at $T_d\sim 20\,\mathrm{K}$ the $850\,\mu$m brightness will vary as 
\begin{equation}\label{equ:4}
F_\mathrm{850}\propto L^{1.5/(4+\beta)},
\end{equation}
where $\beta$ is the dust emissivity index. For JCMTPP\_J182023.2, the SED fitting procedure return $\beta=2$ for the best-fit greybody fit. Using this value for $\beta$, the $850\,\mu$m brightness of JCMTPP\_J182023.2 varies as $F_\mathrm{850}\propto L_{acc}^{0.25}$. For small fractional changes in submillimeter brightness, the rate of change of accretion luminosity is approximately given by
\begin{equation}
\frac{\Delta L_{acc}}{L_{acc}}\propto4\times \frac{\Delta F_\mathrm{850}}{F_\mathrm{850}}. 
\end{equation}
The $850\,\mu$m flux of JCMTPP\_J182023.2 increased at a level of $4\%$ over 3.5 yrs therefore implies a $\sim16\%$ rise in $L_{acc}$ or accretion rate of the central protostar. 

The expected burst of submillimeter flux of JCMTPP\_J182023.2 is two orders of magnitude lower than the submillimeter burst of the two outbursting MYSOs NGC\,6334I\,MM1 and S255IR NIRS3, whose submillimeter luminosity bursts were measured to be $70\pm20$ and $\sim16$ from interferometric observations at around $850\,\mu$m in two epochs \citep{2017ApJ...837L..29H,2018ApJ...863L..12L}, respectively. The areas used for measuring the submillimeter bursts of NGC\,6334I\,MM1 and S255IR NIRS3 are on the order of $1\arcsec$, much smaller than the beam size ($14\farcs4$) of JCMT-SCUBA2 at $850\,\mu$m. The closer to the outbursting central protostar, the response in dust temperature may be stronger. Moreover, the dust temperature in the envelope of protostar, is the result of balance between central source heating and interstellar radiation field heating. Especially for the high-mass star-forming complex where the interstellar radiation field is strong \citep{2006A&A...449..609J}, temperature changes due to accretion luminosity variability may be muted by this external radiation component. 

Thus, in M\,17, the strong interstellar radiation field might significantly weaken the change of dust temperature in the envelope due to the luminosity outburst of the central protostar. For JCMTPP\_J182023.2, a source in the M\,17 field, its luminosity outburst of $\sim16\%$ scaled from its submillimeter peak flux change (4\%) likely represents a lower limit for the luminosity burst of the central embedded protostar.

\begin{figure}
\centering
\includegraphics[width=0.45\textwidth]{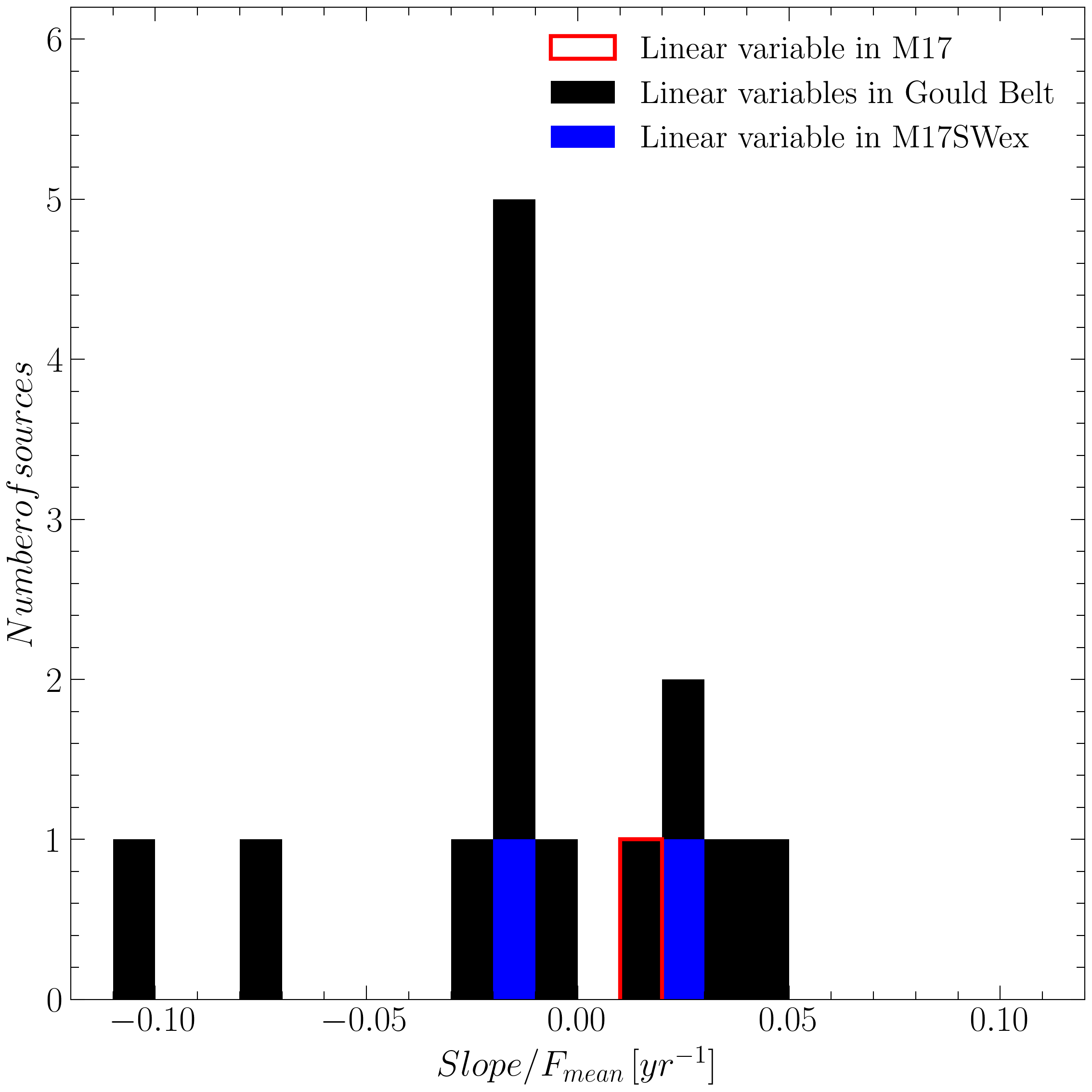}
\caption{Histograms of the fractional slopes determined by linear fit. The black and blue histogram shows the slopes of linear variables in the Gould belt and M17\,SWex recovered by the JCMT Transient survey, respectively. }
\label{fig:850_stats}
\end{figure}

\subsection{Comparison with other JCMT Transient Survey fields}\label{sect:comparison}
Initially, eight fields from the Gould Belt were monitored by the JCMT Transient Survey \citep{2017ApJ...849...43H}. 
After four years of observations, \citet{2021ApJ...920..119L} searched for and characterized variability on 295 submillimeter peaks brighter than $0.14\,\mathrm{Jy\,beam^{-1}}$ at $850\,\mu$m from these fields. Eighteen out of 83 protostars in the sample were found to exhibit secular variability at submillimeter wavelength \citep{2021ApJ...920..119L}. Increasing the monitoring time to six years, 18 secular variables were robustly recovered along with an additional 20 candidate variables \citep{2024ApJ...966..215M}. This underscores the importance of long-term monitoring to uncover secular variability.

In M\,17, we identified only one secular variable candidate, out of 164 peaks with $F_\mathrm{850} > 0.14\,\mathrm{Jy\,beam^{-1}}$. \citet{2024AJ....168..122P} identified two secular variable candidates out of 146 peaks with $F_\mathrm{850} > 0.1\,\mathrm{Jy\,beam^{-1}}$ in M17\,SWex. This corresponds to a lower detection rate than the 18 variables out of 83 protostars within the eight fields of the Gould Belt found by \citet{2021ApJ...920..119L} over a similar timescale. 
Moreover, the secular variable JCMTPP\_J182023.2 varies by only $1.1\%/\mathrm{yr}$ over the 3.5-year monitoring for M\,17. With a similar timescale of 4 years, the JCMT monitoring for the Gould Belt uncovered 14 linear variables \citep{2021ApJ...920..119L} and two linear variables for M17\,SWex \citep{2024AJ....168..122P}.
Figure~\ref{fig:850_stats} compares the fractional change per year in $F_\mathrm{850}$ of the 16 robust linear variables in Gould Belt and M17\,SWex with that of JCMTPP\_J182023.2 in M\,17. The absolute linear slope of JCMTPP\_J182023.2 is close to the lowest value ($0.96\%$) of the absolute slopes of the 14 robust linear variables found in Gould Belt, but is comparable to the slopes of the two linear variables in M17\,SWex at a similar distance. 

M\,17 is at least four times farther away than the eight fields of Gould Belt within 500\,pc. It is worth recognizing challenges of detecting submillimeter variability signatures at much greater distances, where for a fixed JCMT-SCUBA2 beam ($\mathrm{HPBW}\sim14\farcs4$ at $850\,\mu$m) the amount of envelope and nearby cloud contributing to the observed peak flux increases significantly and may therefore smooth over the localized variations. As discussed in Section~\ref{sect:lumin}, the interstellar radiation field might also contribute to the dust temperature in the envelope of a protostar. We found a dust temperature of 20 K for the envelope of JCMTPP\_J182023.2 returned by the greybody fit. In contrast, the northern condensation in M17\,SW, which is much closer than JCMTPP\_J182023.2 to the M\,17 \ion{H}{2} region, was found to have a dust temperature of 30 K derived also from the graybody fit \citep{1993MNRAS.264.1025H}. The closer to the \ion{H}{2} region, the higher dust temperature is found. The strong interstellar radiation field from the M\,17 \ion{H}{2} region might somewhat lower the peak flux variability of the submillimeter sources distributed around the \ion{H}{2} region, through its contribution to the envelope temperature \citep{2006A&A...449..609J}. Thus, combined, the greater distance and stronger interstellar radiation likely combine to reduce the detection rate of secular variables at submillimeter wavelengths for the high-mass star-forming region M\,17. An in-depth comparison between distant regions (M\,17 complex, S255 and DR21) and the Gould Belt regions will be presented in a separate paper (Wang et al. in prepration). The JCMT Transient survey team is also preparing a summary paper that includes the comparison between the regions observed by the survey.

Our mid-IR analyses of M\,17 made use of the WISE/NEOWISE monitoring data from the period of 2010 to 2022 December. From the same WISE/NEOWISE data sets, \citet{2024AJ....168..122P} classified 41 variables at W2 in M17\,SWex using the same methods as for M\,17. The sample of 41 candidate variable YSOs in M17\,SWex almost doubles the sample of 22 candidate variable YSOs classified in M\,17. In M17\,SWex,  $56.1\%$ of the sample are Class I YSOs \citep{2024AJ....168..122P}, however, a somewhat lower fraction, 38.1\% of the sample in M\,17 are Class I YSOs. This difference between the two regions at similar distances might be attributed to the sequential star formation of the M\,17 star-forming complex, with the M\,17 \ion{H}{2} region older than the infrared dark cloud M17\,SWex \citep{2009ApJ...696.1278P,2010ApJ...714L.285P,2016ApJ...825..125P}. For mid-IR variability of the candidate YSOs in M\,17, we found consistency with those in M17\,SWex \citep{2024AJ....168..122P} in the relative numbers of variables by evolutionary class and variability type, with a higher fraction of Class I than Class II being variable. However, the overall fraction of variability at both evolutionary stages in M\,17 and in M17\,SWex \citep{2024AJ....168..122P} are notably lower than those in the Gould Belt \citep{2021ApJ...920..132P}. Such a difference does not mean a higher fraction of YSOs with variability in the Gould Belt than in M\,17 and M17\,SWex. In fact, only 40\% of the initial YSO sample in M\,17 (30\% for M17\,SWex; \citealt{2024AJ....168..122P}) have high-quality NEOWISE data, in contrast this fraction increases to 80\% for the YSO sample in the Gould Belt. The YSO samples in M\,17 and M17\,SWex are likely more biased towards higher mass and/or later stages, lower mass YSOs at early stages probably escape detection even in the infrared because of the high extinction and large distances of the massive star formation clouds. However, the low completeness of the variable YSOs found in M\,17 and M17\,SWex hampers a solid comparison of the mass-dependent incidence of mid-IR variability to those of the Gould Belt.

\section{Summary}\label{sect:sum}
This study presents the first comprehensive attempt to characterize YSO variability at mid-IR and submillimeter wavelengths for the massive star-forming region M\,17. 

We analyze the 3.5 years JCMT Transient Survey data of monthly cadence and the 9 years \textit{NEOWISE} W2 data with a cadence of six months. A candidate linear variable at $850\,\mu$m (JCMTPP\_J182023.2) is identified, with its fractional slope about 8 times the fractional slope uncertainty. The mid-IR analysis reveals significant variability for 22 YSOs. Consistent with the results from other fields within the JCMT Transient survey, eight Gould Belt fields and M17SWex, a higher fraction of Class I YSOs are variable at mid-IR than Class II YSOs, suggesting increased variability at earlier stages of star formation.

The crossmatch between the JCMT submillimeter sources and mid-IR variables returns seven YSOs with JCMT submillimeter sources, three (M17~MIR, SPICY\,81957 and 82001) of which are located within the M\,17 \ion{H}{2} region. Although the three YSOs exhibit significant mid-IR variability, the corresponding submillimeter sources do not show any observable variability at either $450$ or $850\,\mu$m. The candidate linear variable source JCMTPP\_J182023.2 is only visible at wavelengths $\lambda\gtrsim 70\,\mu$m, suggesting that the embedded protostar(s) is (are) in a very early stage. The total mass ($172\,M_\sun$) and bolometric luminosity ($698\,L_\sun$) of JCMTPP\_J182023.2 suggests that the source is in the phase of converting mass into luminosity, in line with the evolutionary path of massive star formation. Compared against other MYSOs exhibiting drastic submillimeter bursts observed with interferometric arrays, the submillimeter flux variation of JCMTPP\_J182023.2 if modest, a $4\%$ increase in 3.5 years at $850\,\mu$m. This brightness change suggests a lower limit of $\sim16\%$ for the potential luminosity burst of JCMTPP\_J182023.2. 

Compared with previous submillimeter and mid-IR variability analyses in the eight nearby Gould Belt regions, we find an order of magnitude decrease in detectable variability within M\,17, which is even lower than that within M17\,SWex, a more quiescent region at a similar distance.  We expect that observing inherent submillimeter and mid-IR variability is more difficult for sources at larger distances and in more complex environments. Our study underscores the necessity of long-term, multi-wavelength monitoring campaigns across a variety of star formation environments to unravel the intricate processes governing the early evolution of YSOs.

\begin{acknowledgments}
We sincerely appreciate the constructive and insightful comments from the anonymous reviewer. This work is supported by the National Natural Science Foundation of China through Grant 1237303. ZC acknowledges the Natural Science Foundation of Jiangsu Province (grant No. BK20231509). GJH is supported by the National Key R\&D program
of China 2022YFA1603102 and by the National Natural Science Foundation of China through grant 12173003. 
DJ is supported by NRC Canada and by an NSERC Discovery Grant.
We acknowledge J.H. for her contribution to the JCMT Transient Survey. This paper makes use of the following ALMA data: ADS/JAO.ALMA\#2018.1.01091.S. ALMA is a partnership of ESO (representing its member states), NSF (USA) and NINS (Japan), together with NRC (Canada), MOST and ASIAA (Taiwan, China), and KASI (Republic of Korea), in cooperation with the Republic of Chile. The Joint ALMA Observatory is operated by ESO, AUI/NRAO and NAOJ.
\end{acknowledgments}

%

\vspace{5mm}
\facilities{James Clark Maxwell Telescope (JCMT), Atacama Large Millimeter/submillimeter Array (ALMA)}


\software{astropy \citep{2013A&A...558A..33A,2018AJ....156..123A}, STARLINK \citep{2014ASPC..485..391C}, TOPCAT \citep{2005ASPC..347...29T}}



\bibliography{myrefs}{}
\bibliographystyle{aasjournal}

\appendix

For completeness, the locations, peak brightnesses, and variability measures used in this paper for all the bright sources found by the Fellwalker algorithm at $450$ and $850\,\mu$m are presented in Tables~\ref{tab:450_sources} and \ref{tab:850_sources}, respectively. In Table~\ref{tab:mir_variable} we list the mid-IR variability measures of the 22 YSOs identified as exhibiting robust mid-IR variability across the six investigated types. 

\begin{deluxetable}{ccccc}[!h]
    \centering
    \footnotesize
    \tablecaption{JCMT Transient Statistics of Bright Sources ($>0.65\,\mathrm{Jy\,beam^{-1}}$ at $450\,\mu$m) in M\,17 \label{tab:450_sources}}
\tablehead{
\colhead{Index} & \colhead{ID} & \colhead{$\langle F_{450}\rangle $} & \colhead{$\mathrm{SD}$} & \colhead{$\mathrm{SD/SD_{fid}}$}  \\
      &    & \colhead{($\mathrm{mJy\,bm^{-1}}$)} &   \colhead{($\mathrm{mJy\,bm^{-1}}$)}   &   }
\startdata
1 & JCMTPP\_J182022.4-161128 & 73765 & 1375 & 0.7 \\
2 & JCMTPP\_J182022.7-161156 & 70575 & 1161 & 0.6 \\
3 & JCMTPP\_J182023.8-161140 & 67712 & 1451 & 0.8 \\
4 & JCMTPP\_J182021.1-161234 & 66886 & 1492 & 0.8 \\
5 & JCMTPP\_J182021.5-161252 & 61609 & 2184 & 1.3 \\
6 & JCMTPP\_J182022.9-161230 & 50367 & 1264 & 0.9 \\
7 & JCMTPP\_J182018.8-161126 & 43181 & 936 & 0.8 \\
8 & JCMTPP\_J182024.3-161326 & 43425 & 1217 & 1.0 \\
9 & JCMTPP\_J182022.6-161258 & 42492 & 1417 & 1.2 \\
10 & JCMTPP\_J182019.4-161228 & 40712 & 1362 & 1.2 \\
11 & JCMTPP\_J182023.6-161254 & 38922 & 1300 & 1.2 \\
12 & JCMTPP\_J182025.1-161350 & 31763 & 791 & 0.9 \\
\bf 13 & JCMTPP\_J182023.3-161448 & 30173 & 566 & 0.7 
\enddata
\tablecomments{Table~\ref{tab:450_sources} is published in its entirety in the electronic edition. A portion is shown here for guidance on its form and content.}
\end{deluxetable}


\begin{deluxetable}{ccccc}[!h]
    \centering
    \footnotesize
    \tablecaption{JCMT Transient Statistics of Bright Sources ($>0.14\,\mathrm{Jy\,beam^{-1}}$ at $850\,\mu$m) in M\,17 \label{tab:850_sources}}
\tablehead{
\colhead{Index} & \colhead{ID} & \colhead{$\langle F_{850}\rangle $} & \colhead{$\mathrm{SD}$} & \colhead{$\mathrm{SD/SD_{fid}}$}  \\
      &    & \colhead{($\mathrm{mJy\,bm^{-1}}$)} &   \colhead{($\mathrm{mJy\,bm^{-1}}$)}   &     }
\startdata
1 & JCMTPP\_J182022.4-161127 & 17954 & 181 & 0.8 \\
2 & JCMTPP\_J182022.8-161154 & 15409 & 180 & 0.9 \\
3 & JCMTPP\_J182023.6-161139 & 14873 & 233 & 1.2 \\
4 & JCMTPP\_J182022.8-161012 & 929 & 31 & 1.6 \\
5 & JCMTPP\_J182030.7-161051 & 546 & 17 & 1.0 \\
6 & JCMTPP\_J182035.5-161127 & 311 & 15 & 1.0 \\
7 & JCMTPP\_J182041.5-161148 & 229 & 18 & 1.2 \\
8 & JCMTPP\_J182021.1-161236 & 13166 & 166 & 1.0 \\
9 & JCMTPP\_J182022.6-161230 & 10057 & 185 & 1.4 \\
10 & JCMTPP\_J182024.3-161324 & 8798 & 81 & 0.7 \\
\bf 112 & JCMTPP\_J182023.2-161448 & 5831 & 79 & 1.0  
\enddata
\tablecomments{Table~\ref{tab:850_sources} is published in its entirety in the electronic edition. A portion is shown here for guidance on its form and content.}
\end{deluxetable}



\begin{deluxetable*}{ccccccccccccc}
    \centering
    \footnotesize
    \tablecaption{Variability Statistics of YSO Candidates in M\,17 \label{tab:mir_variable}}
\tablehead{
\colhead{\#} & \colhead{Alias} & \colhead{R.A. (deg)} & \colhead{Decl. (deg)} & \colhead{$\mathrm{N_{W2}}$} & \colhead{Class} & \colhead{$\mathrm{SD}/\sigma$} & \colhead{$\Delta \mathrm{W2}$} & \colhead{$\mathrm{FAP_{LSP}}$} & \colhead{Period} & \colhead{$\mathrm{FAP_{LIN}}$} & \colhead{Slope}  & \colhead{Type}  \\
&                            & \colhead{(J2000)}    &  \colhead{(J2000)}    &                              &                & 
                               & \colhead{(mag)}                &                                & \colhead{(day)}  &                 &\colhead{(\%/yr)} &                 }
\startdata
1 & SPICY\,81172 & 274.80060 & -16.16145 & 19 &  II & 1.97 & 0.26 & 4.83e-03 & 2957.9 & 7.32e-02 & -3.15 & curved \\
2 & SPICY\,81351 & 274.84851 & -16.30373 & 19 &  II & 4.08 & 1.10 & - & - & - & - & irregular \\
3 & SPICY\,81362 & 274.85173 & -16.12506 & 19 &  II & 2.27 & 0.31 & 4.13e-05 & 3660.3 & 1.59e-04 & 5.59 & curved \\
4 & SPICY\,81397 & 274.86728 & -16.06855 & 19 &  II & 0.99 & 0.17 & 6.02e-03 & 3660.3 & 4.12e-03 & 1.97 & curved \\
5 & SPICY\,81623 & 274.96089 & -15.91883 & 19 &  I & 3.76 & 0.75 & 2.46e-03 & 4800.0 & 2.81e-03 & -7.22 & curved \\
6 & SPICY\,81642 & 274.97296 & -15.93737 & 19 & un. & 1.63 & 0.66 & 1.15e-03 & 437.8 & 6.70e-01 & 0.98 & periodic \\
7 & SPICY\,81671 & 274.98976 & -16.20139 & 19 &  II & 2.29 & 0.34 & 2.17e-06 & 4800.0 & 3.03e-06 & 6.17 & linear \\
8 & SPICY\,81681 & 274.99467 & -16.19189 & 19 &  I & 3.41 & 2.37 & 1.40e-04 & 247.7 & 9.98e-01 & -0.12 & periodic \\
9 & SPICY\,81687 & 274.99663 & -16.03847 & 19 &  I & 2.25 & 0.67 & - & - & - & - & burst \\
10 & SPICY\,81696 & 275.00495 & -16.04284 & 19 &  I & 1.01 & 0.18 & 3.75e-03 & 2481.7 & 8.01e-01 & 0.45 & periodic \\
11 & SPICY\,81724 & 275.01634 & -16.04341 & 19 & un. & 1.48 & 0.44 & 1.96e-03 & 4800.0 & 1.12e-04 & -3.75 & curved \\
12 & SPICY\,81777 & 275.03838 & -16.06545 & 19 &  II & 5.01 & 1.36 & 4.74e-04 & 232.8 & 3.48e-01 & -4.58 & periodic \\
13 & SPICY\,81896 & 275.10147 & -15.94373 & 19 &  II & 2.20 & 0.59 & - & - & - & - & burst \\
14 & SPICY\,81925 & 275.11978 & -16.08311 & 19 &  II & 2.01 & 0.75 & - & - & - & - & burst \\
15 & SPICY\,81957 & 275.13680 & -16.02584 & 19 &  I & 1.70 & 0.89 & 7.85e-04 & 2957.7 & 9.67e-05 & 5.51 & linear \\
16 & SPICY\,81974 & 275.14552 & -16.02265 & 19 &  I & 1.17 & 0.25 & 2.29e-03 & 4800.0 & 3.51e-01 & 1.11 & curved \\
17 & SPICY\,81998 & 275.15714 & -16.33176 & 19 &  II & 1.54 & 0.62 & 9.00e-03 & 383.3 & 2.43e-02 & -2.19 & periodic \\
18 & SPICY\,82001 & 275.15779 & -16.02748 & 13 & un. & 3.90 & 1.01 & - & - & - & - & irregular \\
19 & SPICY\,82007 & 275.16341 & -15.99153 & 19 &  II & 5.86 & 0.95 & 2.47e-04 & 403.3 & 2.71e-02 & -10.11 & periodic \\
20 & SPICY\,82165 & 275.25569 & -15.99572 & 19 &  II & 3.16 & 0.50 & - &- & - &  - & irregular \\
21 & SPICY\,82315 & 275.33538 & -16.12251 & 19 &  I & 3.28 & 1.00 & - & - & - & - & irregular \\
22 & M17\,MIR & 275.09586 & -16.19659 & 20 &  I & 6.69 & 2.13 & - & - & 8.28e-06 & 7.32 & irregular \\
\enddata
\tablenotetext{}{}
\end{deluxetable*}

%




\end{CJK*}
\end{document}